\documentclass[sigconf]{acmart}

\AtBeginDocument{%
  \providecommand\BibTeX{{%
    \normalfont B\kern-0.5em{\scshape i\kern-0.25em b}\kern-0.8em\TeX}}}

\usepackage[utf8]{inputenc}
\usepackage{xcolor}
\usepackage{amsmath}

\usepackage[acronyms,nonumberlist,nopostdot,nomain,nogroupskip]{glossaries}
\usepackage{booktabs}

\usepackage{tabularx}

\usepackage{tikz}
\usepackage{pgfplots}
\pgfplotsset{compat=newest}
\pgfplotsset{plot coordinates/math parser=false}
\newlength\fheight
\newlength\fwidth
\usetikzlibrary{plotmarks,patterns,decorations.pathreplacing,backgrounds,calc,arrows,arrows.meta,spy,matrix}
\usepgfplotslibrary{patchplots,groupplots,colorbrewer}
\usepackage{tikzscale}

\newif\ifexttikz
\exttikzfalse

\ifexttikz
	\usetikzlibrary{external}
\fi

\usepackage{multirow}
\usepackage[font=footnotesize]{subcaption}
\usepackage[font=footnotesize]{caption}

\usepackage{mathtools}

\newacronym{6g}{6G}{sixth generation}
\newacronym{3gpp}{3GPP}{3rd Generation Partnership Project}
\newacronym{adc}{ADC}{Analog to Digital Converter}
\newacronym{dac}{DAC}{Digital to Analog Converter}
\newacronym{5g}{5G}{5th generation}
\newacronym{aimd}{AIMD}{Additive Increase Multiplicative Decrease}
\newacronym{am}{AM}{Acknowledged Mode}
\newacronym{amc}{AMC}{Adaptive Modulation and Coding}
\newacronym{aoa}{AoA}{Angle of Arrival}
\newacronym{aod}{AoD}{Angle of Departure}
\newacronym{aqm}{AQM}{Active Queue Management}
\newacronym{awgn}{AGWN}{Additive White Gaussian Noise}
\newacronym{balia}{BALIA}{Balanced Link Adaptation}
\newacronym{bdp}{BDP}{Bandwidth-Delay Product}
\newacronym{bf}{BF}{Beamforming}
\newacronym{fpga}{FPGA}{field-programmable gate array}
\newacronym{cc}{CC}{Congestion Control}
\newacronym{cdf}{CDF}{Cumulative Distribution Function}
\newacronym{cn}{CN}{Core Network}
\newacronym{cm}{CM}{confusion matrix}
\newacronym[plural=\gls{cnn}s,firstplural=convolutional neural networks (CNNs)]{cnn}{CNN}{convolutional neural network}
\newacronym{cqi}{CQI}{Channel Quality Information}
\newacronym{cp}{CP}{Control Plane}
\newacronym{csirs}{CSI-RS}{Channel State Information - Reference Signal}
\newacronym{dc}{DC}{Dual Connectivity}
\newacronym{dce}{DCE}{Direct Code Execution}
\newacronym{dci}{DCI}{Downlink Control Information}
\newacronym{dmr}{DMR}{Deadline Miss Ratio}
\newacronym{dmrs}{DMRS}{DeModulation Reference Signal}
\newacronym{e2e}{E2E}{End-to-End}
\newacronym{ecn}{ECN}{Explicit Congestion Notification}
\newacronym{ebs}{EBS}{exhaustive beam sweep}
\newacronym{edf}{EDF}{Earliest Deadline First}
\newacronym{enb}{eNB}{evolved Node Base}
\newacronym{epc}{EPC}{Evolved Packet Core}
\newacronym{es}{ES}{Edge Server}
\newacronym{fdma}{FDMA}{Frequency Division Multiple Access}
\newacronym{fdd}{FDD}{Frequency Division Duplexing}
\newacronym[firstplural=Radio Access Technologies (RATs)]{rat}{RAT}{Radio Access Technology}
\newacronym{fs}{FS}{Fast Switching}
\newacronym{txer}{TX}{transmitter}
\newacronym{rxer}{RX}{receiver}
\newacronym{bt}{BT}{beam tracking}
\newacronym{ftp}{FTP}{File Transfer Protocol}
\newacronym{gnb}{gNB}{Next Generation Node Base}
\newacronym{bs}{BS}{Base Station}
\newacronym{harq}{HARQ}{Hybrid Automatic Repeat reQuest}
\newacronym{hetnet}{HetNet}{Heterogeneous Network}
\newacronym{hh}{HH}{Hard Handover}
\newacronym{hol}{HOL}{Head-of-Line}
\newacronym{ia}{IA}{initial access}
\newacronym{imt}{IMT}{International Mobile Telecommunication}
\newacronym{iot}{IoT}{Internet of Things}
\newacronym{los}{LOS}{Line-of-Sight}
\newacronym{lte}{LTE}{Long Term Evolution}
\newacronym{m2m}{M2M}{Machine to Machine}
\newacronym{ml}{ML}{machine learning}
\newacronym{dl}{DL}{deep learning}
\newacronym{mac}{MAC}{Medium Access Control}
\newacronym{mc}{MC}{Multi-Connectivity}
\newacronym{mcs}{MCS}{Modulation and Coding Scheme}
\newacronym{mec}{MEC}{Mobile Edge Cloud}
\newacronym{mi}{MI}{Mutual Information}
\newacronym{mimo}{MIMO}{Multiple Input, Multiple Output}
\newacronym{mmwave}{mmWave}{millimeter wave}
\newacronym{mmWave}{mmWave}{Millimeter wave}
\newacronym{mptcp}{MPTCP}{Multipath TCP}
\newacronym{mr}{MR}{Maximum Rate}
\newacronym{mss}{MSS}{Maximum Segment Size}
\newacronym{mtd}{MTD}{Machine-Type Device}
\newacronym{mtu}{MTU}{Maximum Transmission Unit}
\newacronym{nfv}{NFV}{Network Function Virtualization}
\newacronym{nlos}{NLOS}{Non-Line-of-Sight}
\newacronym{nr}{NR}{New Radio}
\newacronym{ofdm}{OFDM}{Orthogonal Frequency Division Multiplexing}
\newacronym{pdcch}{PDCCH}{Physical Downlink Control Channel}
\newacronym{pdcp}{PDCP}{Packet Data Convergence Protocol}
\newacronym{pdsch}{PDSCH}{Physical Downlink Shared Channel}
\newacronym{pdu}{PDU}{Packet Data Unit}
\newacronym{pf}{PF}{Proportional Fair}
\newacronym{pgw}{PGW}{Packet Gateway}
\newacronym{phy}{PHY}{Physical}
\newacronym{pbch}{PBCH}{Physical Broadcast Channel}
\newacronym[plural=\gls{mme}s,firstplural=Mobility Management Entities (MMEs)]{mme}{MME}{Mobility Management Entity}
\newacronym{prb}{PRB}{Physical Resource Block}
\newacronym{pss}{PSS}{Primary Synchronization Signal}
\newacronym{pucch}{PUCCH}{Physical Uplink Control Channel}
\newacronym{pusch}{PUSCH}{Physical Uplink Shared Channel}
\newacronym{rach}{RACH}{Random Access Channel}
\newacronym{ran}{RAN}{Radio Access Network}
\newacronym{red}{RED}{Random Early Detection}
\newacronym{rf}{RF}{Radio Frequency}
\newacronym{rlc}{RLC}{Radio Link Control}
\newacronym{rlf}{RLF}{Radio Link Failure}
\newacronym{rrc}{RRC}{Radio Resource Control}
\newacronym{rrm}{RRM}{Radio Resource Management}
\newacronym{rr}{RR}{Round Robin}
\newacronym{rs}{RS}{Remote Server}
\newacronym{rsrp}{RSRP}{Reference Signal Received Power}
\newacronym{rss}{RSS}{Received Signal Strength}
\newacronym{rtt}{RTT}{Round Trip Time}
\newacronym{rw}{RW}{Receive Window}
\newacronym{rx}{RX}{Receiver}
\newacronym{sa}{SA}{standalone}
\newacronym{sack}{SACK}{Selective Acknowledgment}
\newacronym{sap}{SAP}{Service Access Point}
\newacronym{ap}{AP}{Access Point}
\newacronym{sch}{SCH}{Secondary Cell Handover}
\newacronym{scoot}{SCOOT}{Split Cycle Offset Optimization Technique}
\newacronym{sdma}{SDMA}{Spatial Division Multiple Access}
\newacronym{sinr}{SINR}{Signal to Interference plus Noise Ratio}
\newacronym{sm}{SM}{Saturation Mode}
\newacronym{snr}{SNR}{Signal-to-Noise-Ratio}
\newacronym{son}{SON}{Self-Organizing Network}
\newacronym{ss}{SS}{Synchronization Signal}
\newacronym{ssbs}{SSBs}{synchronization signal blocks}
\newacronym{ssb}{SSB}{synchronization signal block}
\newacronym{srs}{SRS}{Sounding Reference Signal}
\newacronym{sss}{SSS}{Secondary Synchronization Signal}
\newacronym{tb}{TB}{Transport Block}
\newacronym{tcp}{TCP}{Transmission Control Protocol}
\newacronym{tdd}{TDD}{Time Division Duplexing}
\newacronym{tdma}{TDMA}{Time Division Multiple Access}
\newacronym{tfl}{TfL}{Transport for London}
\newacronym{tm}{TM}{Transparent Mode}
\newacronym{trp}{TRP}{Transmitter Receiver Pair}
\newacronym{tti}{TTI}{Transmission Time Interval}
\newacronym{ttt}{TTT}{Time-to-Trigger}
\newacronym{tx}{TX}{Transmitter}
\newacronym{ue}{UE}{User Equipment}
\newacronym{ul}{UL}{Uplink}
\newacronym{uml}{UML}{Unified Modeling Language}
\newacronym{um}{UM}{Unacknowledged Mode}
\newacronym{utc}{UTC}{Urban Traffic Control}
\newacronym{vm}{VM}{Virtual Machine}
\newacronym{rsrq}{RSRQ}{Reference Signal Received Quality}
\newacronym{rssi}{RSSI}{Received Signal Strength Indicator}
\newacronym{crs}{CRS}{Cell Reference Signal}
\newacronym{nsa}{NSA}{Non Stand Alone}
\newacronym{mrdc}{MR-DC}{Multi \gls{rat} \gls{dc}}
\newacronym{endc}{EN-DC}{E-UTRAN-\gls{nr} \gls{dc}}
\newacronym{5gc}{5GC}{5G Core}
\newacronym{si}{SI}{Study Item}
\newacronym{iab}{IAB}{Integrated Access and Backhaul}
\newacronym{wf}{WF}{Wired-first}
\newacronym{hqf}{HQF}{Highest-quality-first}
\newacronym{pa}{PA}{Position-aware}
\newacronym{mlr}{MLR}{Maximum-local-rate}
\newacronym{wbf}{WBF}{Wired Bias Function}
\newacronym{mib}{MIB}{Master Information Block}
\newacronym{sib}{SIB}{Secondary Information Block}
\newacronym{kpi}{KPI}{Key Performance Indicator}
\newacronym{ppp}{PPP}{Poisson Point Process}
\newacronym{gtp}{GTP}{GPRS Tunneling Protocol}
\newacronym{amf}{AMF}{Access and Mobility Management Function}
\newacronym{dash}{DASH}{Dynamic Adaptive Streaming over HTTP}
\newacronym{http}{HTTP}{HyperText Transfer Protocol}
\newacronym{qos}{QoS}{Quality of Service}
\newacronym{udp}{UDP}{User Datagram Protocol}
\newacronym{cu}{CU}{Central Unit}
\newacronym{du}{DU}{Distributed Unit}
\newacronym{mt}{MT}{Mobile Termination}
\newacronym{sdap}{SDAP}{Service Data Adaptation Protocol}
\newacronym{tdm}{TDM}{Time Division Multiplexing}
\newacronym{fdm}{FDM}{Frequency Division Multiplexing}
\newacronym{sdm}{SDM}{Space Division Multiplexing}
\newacronym{dag}{DAG}{Directed Acyclic Graph}
\newacronym{st}{ST}{Spanning Tree}
\newacronym{ummimo}{UM-MIMO}{Ultra-massive Multiple Input, Multiple Output}
\newacronym{uavs}{UAVs}{Unmanned Aerial Vehicles}
\newacronym{wlan}{WLAN}{Wireless LAN}
\newacronym{rlnc}{RLNC}{Random Linear Network Coding}
\newacronym{drx}{DRX}{Discontinuous Reception}
\newacronym{cpu}{CPU}{Central Processing Unit}
\newacronym{txb}{TXB}{transmitter's beam}
\newacronym{rxb}{RXB}{receiver's beam}
\newacronym{sifs}{SIFS}{Short Interframe Space}
\newacronym{difs}{DIFS}{DCF Interframe Space}
\tikzstyle{startstop} = [rectangle, rounded corners, minimum width=2cm, minimum height=0.5cm,text centered, draw=black]
\tikzstyle{io} = [trapezium, trapezium left angle=70, trapezium right angle=110, minimum width=3cm, minimum height=1cm, text centered, draw=black]
\tikzstyle{process} = [rectangle, minimum width=2cm, minimum height=0.5cm, text centered, draw=black, alignb=center]
\tikzstyle{decision} = [ellipse, minimum width=2cm, minimum height=1cm, text centered, draw=black]
\tikzstyle{arrow} = [thick,<->,>=stealth]
\tikzstyle{line} = [thick,>=stealth]
\tikzstyle{darrow} = [thick,<->,>=stealth,dashed]
\tikzstyle{sarrow} = [thick,->,>=stealth]
\tikzstyle{larrow} = [line width=0.1mm,dashdotted,->,>=stealth]

\makeatletter
\def\grd@save@target#1{%
  \def\grd@target{#1}}
\def\grd@save@start#1{%
  \def\grd@start{#1}}
\tikzset{
  grid with coordinates/.style={
    to path={%
      \pgfextra{%
        \edef\grd@@target{(\tikztotarget)}%
        \tikz@scan@one@point\grd@save@target\grd@@target\relax
        \edef\grd@@start{(\tikztostart)}%
        \tikz@scan@one@point\grd@save@start\grd@@start\relax
        \draw[minor help lines] (\tikztostart) grid (\tikztotarget);
        \draw[major help lines] (\tikztostart) grid (\tikztotarget);
        \grd@start
        \pgfmathsetmacro{\grd@xa}{\the\pgf@x/1cm}
        \pgfmathsetmacro{\grd@ya}{\the\pgf@y/1cm}
        \grd@target
        \pgfmathsetmacro{\grd@xb}{\the\pgf@x/1cm}
        \pgfmathsetmacro{\grd@yb}{\the\pgf@y/1cm}
        \pgfmathsetmacro{\grd@xc}{\grd@xa + \pgfkeysvalueof{/tikz/grid with coordinates/major step x}}
        \pgfmathsetmacro{\grd@yc}{\grd@ya + \pgfkeysvalueof{/tikz/grid with coordinates/major step y}}
        \foreach \x in {\grd@xa,\grd@xc,...,\grd@xb}
        \node[anchor=north] at (\x,\grd@ya) {\pgfmathprintnumber{\x}};
        \foreach \y in {\grd@ya,\grd@yc,...,\grd@yb}
        \node[anchor=east] at (\grd@xa,\y) {\pgfmathprintnumber{\y}};
      }
    }
  },
  minor help lines/.style={
    help lines,
    gray,
    line cap =round,
    xstep=\pgfkeysvalueof{/tikz/grid with coordinates/minor step x},
    ystep=\pgfkeysvalueof{/tikz/grid with coordinates/minor step y}
  },
  major help lines/.style={
    help lines,
    line cap =round,
    line width=\pgfkeysvalueof{/tikz/grid with coordinates/major line width},
    xstep=\pgfkeysvalueof{/tikz/grid with coordinates/major step x},
    ystep=\pgfkeysvalueof{/tikz/grid with coordinates/major step y}
  },
  grid with coordinates/.cd,
  minor step x/.initial=.5,
  minor step y/.initial=.2,
  major step x/.initial=1,
  major step y/.initial=1,
  major line width/.initial=1pt,
}
\makeatother

\definecolor{desireRed}{RGB}{230,57,60}%
\definecolor{darkPurple}{RGB}{59,31,43}%
\definecolor{springGreen}{RGB}{37,223,145}%
\definecolor{queenBlue}{RGB}{69,123,157}%
\definecolor{spaceCadet}{RGB}{29,53,87}%

\copyrightyear{2021}
\acmYear{2021}
\setcopyright{acmlicensed}
\acmConference[MobiHoc '21]{The Twenty-second International Symposium on Theory, Algorithmic Foundations, and Protocol Design for Mobile Networks and Mobile Computing}{July 26--29, 2021}{Shanghai, China}
\acmBooktitle{The Twenty-second International Symposium on Theory, Algorithmic Foundations, and Protocol Design for Mobile Networks and Mobile Computing (MobiHoc '21), July 26--29, 2021, Shanghai, China}
\acmPrice{15.00}
\acmDOI{10.1145/3466772.3467035}
\acmISBN{978-1-4503-8558-9/21/07}

\begin{document}

\title{\textit{DeepBeam:} Deep Waveform Learning for Coordination-Free Beam Management in mmWave Networks}

\author{Michele Polese, Francesco Restuccia, and Tommaso Melodia}
\affiliation{
\institution{Institute for the Wireless Internet of Things, Northeastern University, Boston, MA, United States}
\country{}
}

\flushbottom
\setlength{\parskip}{0ex plus0.1ex}

\glsunset{nr}
\glsunset{lte}

\begin{abstract}
Highly directional \gls{mmwave} radios need to perform \textit{beam management} to establish and maintain reliable links. To achieve this objective, existing solutions mostly rely on explicit coordination between the \gls{txer} and the \gls{rxer}, which significantly reduces the airtime available for communication and further complicates the network protocol design. This paper advances the state of the art by presenting \emph{DeepBeam}, a framework for beam management that does not require pilot sequences from the \gls{txer}, nor any beam sweeping or synchronization from the \gls{rxer}. This is achieved by inferring (i) the \gls{aoa} of the beam and (ii) the actual beam being used by the transmitter through \emph{waveform-level deep learning} on \textit{ongoing transmissions} between the \gls{txer} to other receivers. In this way, the \gls{rxer} can associate \gls{snr} levels to beams without explicit coordination with the \gls{txer}. This is possible because different beam patterns introduce different ``impairments'' to the waveform, which can be subsequently learned by a \gls{cnn}. To demonstrate the generality of DeepBeam, we conduct an extensive experimental data collection campaign where we collect more than 4 TB of \gls{mmwave} waveforms with (i) 4 phased array antennas at 60.48 GHz, (ii) 2 codebooks containing 24 one-dimensional beams and 12 two-dimensional beams; (iii) 3 receiver gains; (iv) 3 different \gls{aoa}s; (v) multiple \gls{txer} and \gls{rxer} locations. Moreover, we collect waveform data with two custom-designed \gls{mmwave} software-defined radios with fully-digital beamforming architectures at 58 GHz. We also implement our learning models in FPGA to evaluate latency performance. Results show that \emph{DeepBeam} (i) achieves accuracy of up to 96\%, 84\% and 77\% with a 5-beam, 12-beam and 24-beam codebook, respectively; (ii) reduces latency by up to 7x with respect to the 5G NR initial beam sweep in a default configuration and with a 12-beam codebook. The waveform dataset and the full \emph{DeepBeam} code repository are publicly available.
\end{abstract}


\maketitle

\glsresetall


\section{Introduction}
\label{sec:intro}

It is well known that mobile devices are now hungrier than ever for gigabit-per-second data rates \cite{EricssonMobility2020}. Thanks to their promise of data rates orders of magnitude higher than sub-6 GHz technologies \cite{akoum2012coverage}, \gls{mmwave} communications lie at the foundation of \gls{5g} networks and beyond \cite{rangan2017potentials,giordani20196g}.
Among others, one of the core challenges in \gls{mmwave} networks is the severely increased path loss with respect to sub-6 GHz frequencies, which implies that highly-directional communications through beamforming are necessary to bring the transmission range back to acceptable levels \cite{dutta2020case}. As a consequence, the \gls{txer} and the \gls{rxer} need to coordinate to select the beam pair that yields the highest beamforming gain. For this reason, \textit{beam management} in \gls{mmwave} networks has attracted tremendous interest from the research community over the last years \cite{wei2017exhaustive,giordani2018tutorial,li2017design,asadi2018fml}.\vspace{-0.2cm}

\subsection*{\textbf{Background and Motivation}}

\begin{figure}[h!]
	\centering
	\includegraphics[width=\columnwidth]{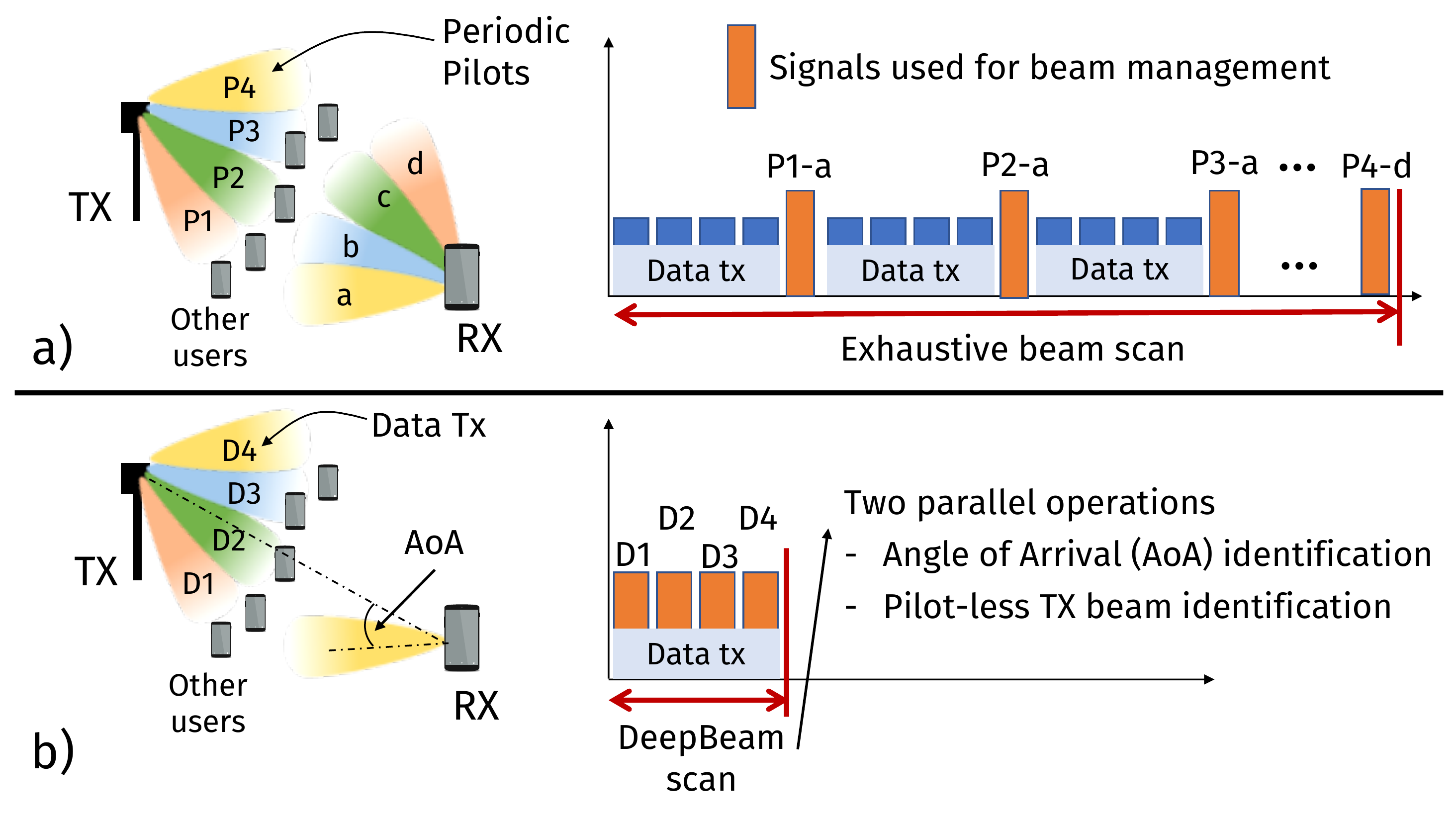}
	\setlength\abovecaptionskip{-.3cm}
	\setlength\belowcaptionskip{-.3cm}
	\caption{Traditional, pilot-based \gls{ebs} (a) vs. DeepBeam (b). In this example, with the \gls{ebs}, the \gls{txer} and \gls{rxer} scan 4 beams each, by transmitting pilots P1-P4 on different \glspl{txb}, and by receiving with beams a-d, respectively, in specific time and frequency resources. With DeepBeam, the \gls{rxer} infers the \gls{aoa} and the \gls{txb} by passively eavesdropping on data transmissions to other users in the network.}
	\label{fig:intuition}
\end{figure}

Beam management is usually a complex procedure that involves several time-consuming steps. First, both the \gls{txer} and the \gls{rxer} need to discover each other by finding the initial beamforming vectors that yield sufficient \gls{snr} to establish a link. This crucial procedure is usually called \textit{initial access} (IA) \glsunset{ia} \cite{giordani2016initial,barati2016initial}. Once the \gls{mmwave} link has been established, \textit{beam tracking} is performed to keep the \gls{txer} and \gls{rxer} beams aligned to avoid sudden drops in \gls{snr}. For both \gls{ia} and beam tracking, the \gls{3gpp} NR standard for \gls{5g} communications utilizes \gls{ssbs}, which are essentially pilot and synchronization sequences that are periodically transmitted by the \gls{txer} in each of its $N_{tx}$ beam directions. By listening on each of its $M_{rx}$ beam directions, the \gls{rxer} is then able to compute the received power for each of the $N_{tx} \cdot M_{rx}$ possible beam combinations, and thus make an informed decision on which beamforming vector to use. The complexity of these beam management techniques, also called \gls{ebs}, is thus quadratic in the number of beams. Figure \ref{fig:intuition}(a) shows an example of \gls{ebs} when $N_{tx} = 4$ and $M_{rx} = 4$ beams are used.
A similar, multi-stage procedure is used for IEEE 802.11ad \cite{ieee80211ad}, where the beams are distributed in $N_{tx} \le 128$ sphere sectors, with beam widths as small as 3 degrees~\cite{Nitsche-infocom2015}. A beam sweep is performed by the \gls{txer} to find the best sector, and, subsequently, intra-sector fine-tuning is used by the \gls{txer} and \gls{rxer} to refine the selection~\cite{nitsche2014ieee}. 

One can intuitively see that pilot-based \gls{ebs} is very inefficient. For example, in 3GPP NR it could take up to 164 milliseconds (ms) to complete an \gls{ia} when 24 beams are used by both \gls{txer} and \gls{rxer} \cite{giordani2018tutorial}. Worse yet, although \gls{ebs} procedures could be feasible in cellular networks, they may not be effective at all in \gls{mmwave} ad hoc networks \cite{zhou2018fastnd}, where links are highly volatile and short in duration.  For this reason, existing work -- discussed in detail in Section \ref{sec:rw} --  has proposed several strategies to improve beam management \cite{Nitsche-infocom2015,va2016beam,DeDonno-ieeetwc2017,palacios2017tracking,Steinmetzer-conext2017,Loch-conext2017,zhou2017beam,sur2018towards,zhou2018beam,Ghasempour-mobicom2018,Haider-mobicom2018,yang2019beam,aykin2020mamba,aykin2020efficient}. However, prior approaches still require coordination between the \gls{txer} and \gls{rxer}, which reduces the effective channel utilization.

In this paper, we take a completely new direction: \textbf{we leverage a data-driven approach based on \glspl{cnn} to achieve coordination-free beam management in \gls{mmwave} networks}. Our core intuition -- explained in details in Section \ref{sec:learning_engine} -- is that different beams will impose different ``impairments'' to the waveform. These impairments will form a unique ``signature'' of the beam, which translates into a small-scale distortion of the  emitted waveform. Therefore, a \gls{cnn} can learn to recognize these unique patterns in the in-phase-quadrature (I/Q) representation of the waveform and thus ultimately distinguish among different beams. The key advantage of our approach is that the beam can be inferred by simply eavesdropping on ongoing transmissions between the \gls{txer} and other \glspl{rxer}, without any explicit synchronization. We use \glspl{cnn} instead of traditional \gls{ml} approaches since they can learn complex features that may not be clearly separable in a low-dimensional space \cite{o2017introduction,OShea-ieeejstsp2018,jian2020deep}, while also meeting real-time constraints \cite{restuccia2019big}. Moreover, \glspl{cnn} allow DeepBeam to be independent from device and scenario through fine-tuning or cross-training techniques, as \glspl{cnn} do not require any antenna- or waveform-specific feature extraction process~\cite{restuccia2020polymorf}.

\subsection*{\textbf{Novel Contributions}}

We summarize the core contributions of this paper below.\smallskip

$\bullet$  We design \emph{DeepBeam}, the first framework that leverages wave\-form-level deep learning to perform beam management in \gls{mmwave} networks without requiring explicit coordination between the \gls{txer} and \gls{rxer} (Section \ref{sec:deepbeam}). Instead, we leverage \textit{ongoing transmissions} between the \gls{txer} to other receivers and infer through \emph{waveform-level deep learning} (i) the \gls{aoa} of the \gls{txb}; and (ii) the \gls{txb} itself. By using these two pieces of information, the \gls{rxer} can infer how to switch its beam toward the \gls{txer}, and can inform the \gls{txer} of which is the best beam to be used for communications with the \gls{rxer}, without the need for explicit pilots. Figure \ref{fig:intuition}(b) summarizes at a very high level why our approach drastically decreases the time taken to identify the best beams to be used for the ongoing \gls{mmwave} links by doing away with pilot-based scanning thanks to deep learning; \smallskip

$\bullet$ Our data-driven approach has been extensively validated with a massive \gls{mmwave} data collection campaign (Section \ref{sec:experimental_setup}). We utilize a well-known experimental \gls{mmwave} prototype by NI \cite{nitestbed} to collect more than 4 TB of \gls{mmwave} waveforms with (i) 4 phased array antennas at 60.48 GHz, (ii) 2 codebooks containing 24 one-dimensional beams (\textit{i.e.}, azimuth only) and 12 two-dimensional beams (\textit{i.e.}, azimuth and elevation); (iii) 3 receiver gains; (iv) 3 different \gls{aoa}s; (v) multiple \gls{txer} and \gls{rxer} locations. Furthermore, we also leverage two custom-designed \gls{mmwave} software-defined radios based on (i) off-the-shelf Xilinx ZCU111 RFSoC-based evaluation boards; and (ii) transceiver boards with 4 fully-digital RF chains, operating in the unlicensed  57-64  GHz frequency  band with  2  GHz  bandwidth~\cite{haarla2020characterizing}. Moreover, we perform a latency analysis of the proposed approach through a \gls{fpga} implementation of our \gls{cnn} (Section~\ref{sec:applications}). Experimental results conclude that DeepBeam \textbf{(i) achieves accuracy of up to 96\%, 84\% and 77\% with a 5-beam, 12-beam and 24-beam codebook, respectively (Section~\ref{sec:results}); (ii) reduces latency by up to 7x with respect to the 5G NR initial beam sweep} in a default configuration and with a 12-beam codebook. We also provide results that investigate how DeepBeam I/Q learning generalizes for training and testing over different phased array antennas and \gls{txer} and \gls{rxer} locations, and investigate cross-training approaches. 
\smallskip

$\bullet$ A major contribution of this paper is that \textbf{we release the experimental waveform dataset to the community, as well as the code used for training and testing our models}\footnote{Dataset: \url{http://hdl.handle.net/2047/D20409451}, repository: \url{https://github.com/wineslab/deepbeam}}. So far, machine learning research in the \gls{mmwave} domain has been severely stymied from the lack of experimental datasets, with most of the current research conducted with data obtained through simulations~\cite{zhou2018deep} and ray-tracing~\cite{alkhateeb2018deep,alkhateeb2019deep,wang2018mmwave}. With our paper, we will enable other researchers to replicate our results and benefit from the data.

\section{The DeepBeam System}\label{sec:deepbeam}

\begin{figure}[t]
    \centering
    \includegraphics[width=.92\columnwidth]{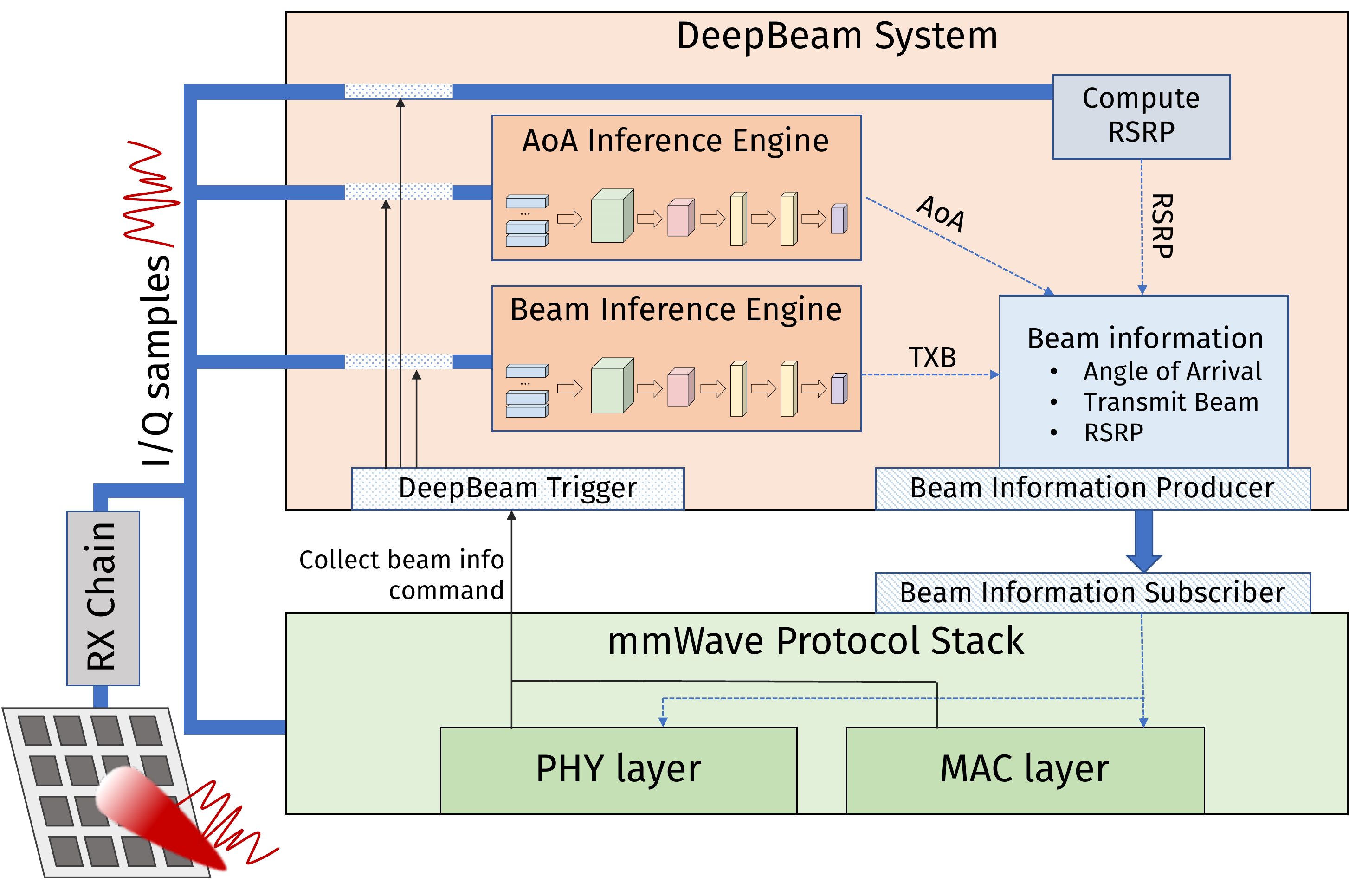}
    \setlength\abovecaptionskip{0.1cm}
    \setlength\belowcaptionskip{-.6cm}
    \caption{DeepBeam Framework. The RX chain converts the analog waveform impinging the phased array into I/Q samples, which are processed by the \gls{phy} and \gls{mac} layers and by DeepBeam. In this, the I/Q samples are used to classify the \gls{txb} and \gls{aoa}, and to estimate the \gls{rsrp}. The mmWave protocol stack connects to DeepBeam through two interfaces, a trigger (to activate DeepBeam) and a data source, to obtain the  \gls{txb}, \gls{aoa}, and \gls{rsrp}.}
    \label{fig:deepbeam}
\end{figure}

The DeepBeam system, shown in Figure~\ref{fig:deepbeam}, is a stand-alone module that can be plugged into the \gls{phy} and \gls{mac} layers of a generic mmWave protocol stack. In other words, it does not rely on any specific feature of, for example, 3GPP NR or IEEE 802.11ad/ay. DeepBeam can be implemented in software, or on \glspl{fpga}, to provide real-time learning with latency guarantees~\cite{restuccia2020deepwierl}. DeepBeam can be easily integrated with the \gls{phy} and \gls{mac} layers through two interfaces, as shown in Figure~\ref{fig:deepbeam}. The first is a trigger that activates DeepBeam when required by the protocol stack. The second is a producer/subscriber interface to which the wireless stack can subscribe to consume the information generated by DeepBeam at its own convenience. This data consists of a list of tuples with three elements, \textit{i.e.}, the \gls{aoa} and the \gls{txb}, both inferred through deep learning, and the associated \gls{rsrp}, a metric that can be used to evaluate the quality of a received signal~\cite{38215}.

The input of DeepBeam consists of the raw digital waveform obtained through the receiver \gls{rf} chain, \textit{i.e.}, the in-phase and quadrature (I/Q) data sampled by the \gls{adc}, without any further processing (\textit{e.g.}, frequency offset tracking, equalization) from the \gls{phy} layer. This means that the module can be directly connected to the device \gls{rf} chain, and that there is no need for synchronization between the transmitter and the receiver, as DeepBeam can handle the I/Q samples even before they are processed at the \gls{phy} layer. Therefore, a device equipped with DeepBeam can passively eavesdrop transmissions in a certain area, and collect statistics on the channel quality associated to the beams that a base station or access point uses to communicate with other users, eventually inferring what is the best beam pair to use for communications.

\begin{figure}[t]
    \centering
    \includegraphics[width=.9\columnwidth]{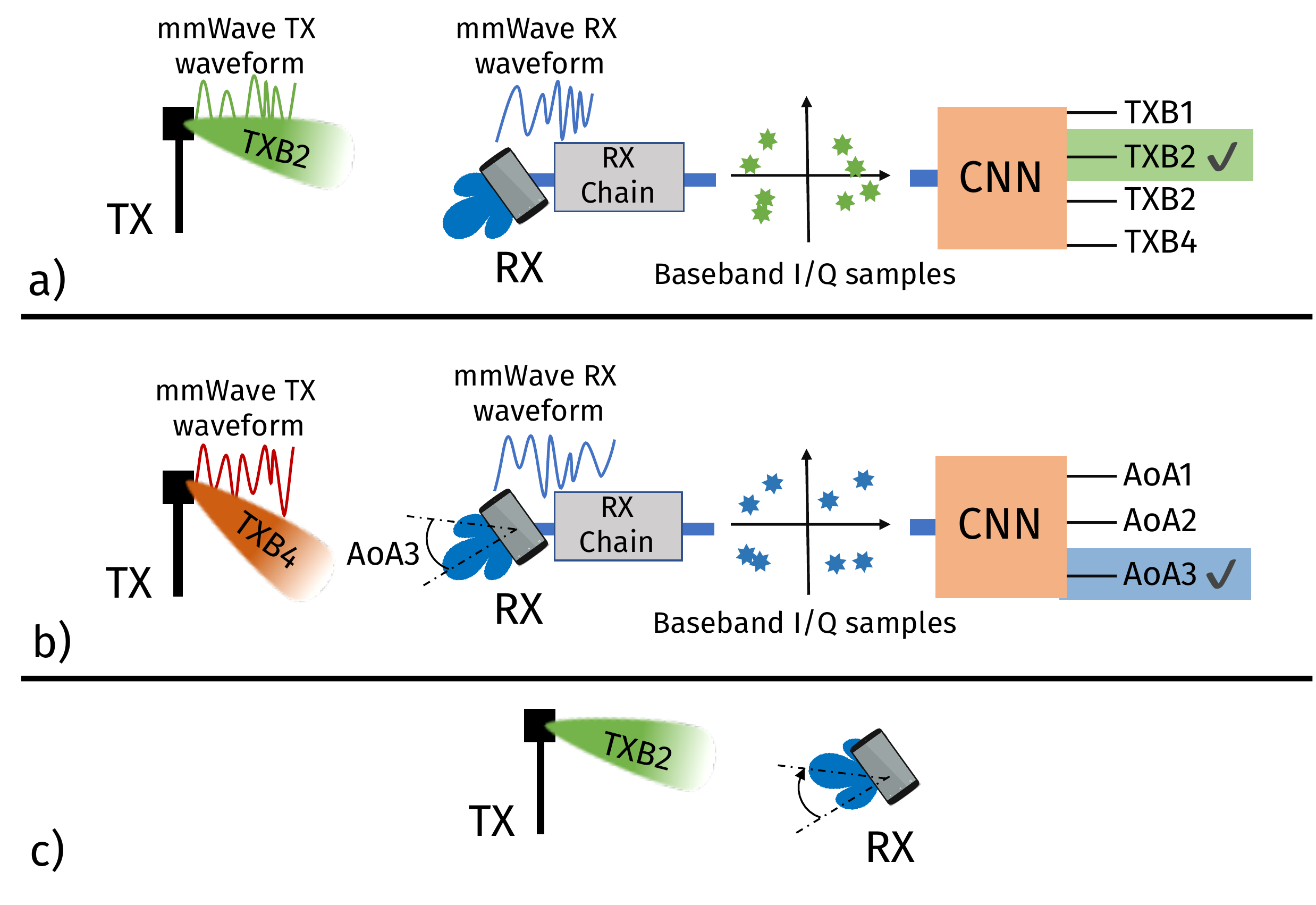}
	\setlength\abovecaptionskip{0cm}
    \setlength\belowcaptionskip{-.5cm}
    \caption{\gls{txb} (a) and \gls{aoa} (b) learning. In (a) and (b), a \gls{cnn} processes the baseband I/Q samples to infer, respectively, the \gls{txb} used by the \gls{txer} and the \gls{aoa}. Then, (c) visually represents the final beam pair selection based on the inferred \gls{txb} and \gls{aoa}.}
    \label{fig:learning}
\end{figure}

The two learning engines at the core of the DeepBeam module are based on \glspl{cnn}, and use the I/Q samples to infer two key elements for beam management procedures, namely the \gls{txb} and \gls{aoa}. These are usually obtained through a pilot-based beam sweep or inference at the transmitter and receiver~\cite{palacios2017tracking,va2016beam}. DeepBeam, instead, can perform the inference on any kind of over-the-air signal, thus speeding up beam management procedures, as we will discuss in Section~\ref{sec:applications}. The \gls{aoa}, as shown in Figure~\ref{fig:learning}, corresponds to the angle with which the received signal impinges the antenna array of the receiver, either through the direct path between the transmitter and the receiver, when in \gls{los}, or through a reflected path, in \gls{nlos}. 
Thanks to this information, the \gls{rxer} can steer the receive beam toward this angular direction to experience the highest beamforming gain. This is shown, as an example, in Figure~\ref{fig:learning}: the \gls{rxer} identifies that the \gls{aoa} is \gls{aoa}3, corresponding to, in this case, $-45^\circ$, and exploits this information to select the matching receive beam. Notice that traditional \gls{aoa} detection methods either require multiple antennas and RF chains~\cite{oumar2012comparison}, or the sampling of the signal in multiple spatial location~\cite{wei2017facilitating}, while DeepBeam can operate directly on I/Q samples from a single RF~chain. \gls{aoa} detection (and \gls{rxer}-beam steering) can improve the performance of DeepBeam itself, as the \gls{rxer} would then be steering toward the \gls{txer} and increase the classification accuracy thanks to a higher \gls{snr}.

DeepBeam also infers through deep learning which beam -- from a certain codebook -- is being used by the \gls{txer} to transmit the waveform just sampled, as shown in Figure~\ref{fig:learning}. Once the DeepBeam-equipped device has eavesdropped on enough ongoing transmissions, it can determine which is the best beam that the \gls{txer} should use to communicate with it, for example, by ranking the inferred \glspl{txb} by the associated \gls{rsrp}. Notice that several state-of-the-art techniques can also be integrated in DeepBeam to minimize the number of transmissions to observe. To this end, we point out that the pilot-less estimation that DeepBeam enables can be considered as a basis for further refinement of fast and efficient beam management schemes. The information regarding the best \gls{txb} can then be used in different ways during the beam management process. For example, in 3GPP NR the mobile device can infer with DeepBeam the best \gls{txb}, and then match it with the next \gls{ssb} in the same angular direction to perform initial access in the proper time and frequency resources. A detailed explanation of how DeepBeam can be used during the 3GPP NR initial access will be given in Section~\ref{sec:applications}, together with an evaluation of the latency reduction obtained by using DeepBeam.

\subsection{Beam and \gls{aoa} Inference Engine}\label{sec:learning_engine}

As mentioned earlier, DeepBeam leverages \glspl{cnn} to perform real-time beam inference. We selected \glspl{cnn} because of their demonstrated performance in addressing complex classification problems in the wireless domain, including modulation classification \cite{OShea-ieeejstsp2018} and radio fingerprinting \cite{riyaz2018deep}. The versatility of \glspl{cnn} is primarily owed to the fact that the filters in the convolutional (Conv) layers learn patterns in the I/Q constellation plane, regardless of where they occur in the waveform (\textit{shift invariance}). This is necessary because there may be rotations in the received signal, induced by different channels, which would otherwise make the framework scenario-dependent. More formally, a Conv layer is trained to learn a series of $F$ filters $\mathbf{P}_f \in \mathbb{R}^{d \times w}, 1 \le f \le F$, where $d$ and $w$ are the depth and width of the filter. An output $\mathbf{O}^f \in \mathbb{R}^{n' \times m'}$ is produced from input $\mathbf{I} \in \mathbb{R}^{n \times m}$ according to the following equation:
\begin{equation}
    \mathbf{O}^{f}_{i,j} = \sum_{k=0}^{d-1}\sum_{\ell=0}^{w-1} \mathbf{P}^f_{d-k, w-\ell} \cdot \mathbf{I}_{i-k, j-\ell}, \, 1 \le i \le n'\ 1\le j \le m',
    \label{eq:filters}
\end{equation}
where $n' = 1 + \left \lfloor{n + d - 2}\right \rfloor$ and $m' = 1 + \left \lfloor{m+w-2}\right \rfloor$. 
This ultimately helps distinguish waveforms far beyond what is possible with traditional dense networks, which were shown to not perform well in RF classification tasks \cite{restuccia2020polymorf}. Furthermore, as  discussed earlier, \glspl{cnn} are amenable to be implemented in \gls{fpga} and integrated in the baseband processing loop \cite{restuccia2019big}, and are easily fine-tunable  \cite{restuccia2020deepwierl}.

\begin{figure}[t]
    \centering
    \includegraphics[width=.87\columnwidth]{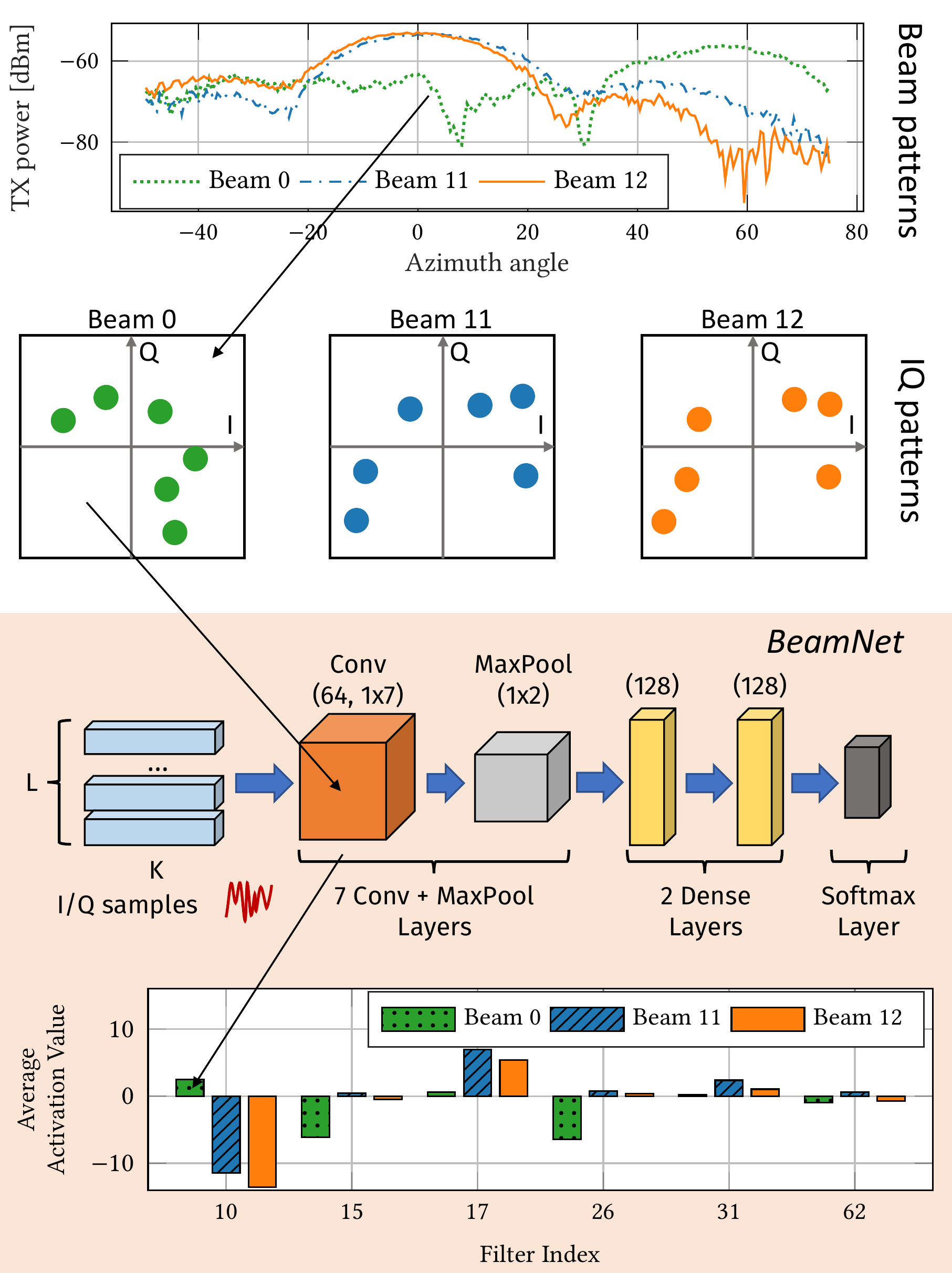}
    \setlength\belowcaptionskip{-.6cm}
    \caption{Top: beam pattern for three different beams (0, 11, 12) of the 24-beams codebook (Sec.~\ref{sec:ni_testbed}). Mid: examples of I/Q patterns for the different beams. Bottom: the baseline architecture of the \textit{BeamNet} \gls{cnn} and average activation value of the first convolutional layer for the same beams. The input of BeamNet is given by $K$ I/Q samples grouped in $L$ blocks.}
    \label{fig:baseline}
\end{figure}

We consider the \emph{BeamNet} \gls{cnn} architecture shown in Figure \ref{fig:baseline}, which we call \emph{baseline}. In the rest of the paper, if not explicitly mentioned otherwise, we will refer to the architecture above. Our baseline has been adapted from the architecture presented in \cite{OShea-ieeejstsp2018}, which has proven to be effective for RF classification tasks. \textit{BeamNet} classifies input tensors of size ($L$, $K$, 2), where $L$ is the number of consecutive input blocks, each composed of $K$ I/Q samples. By increasing the number of  blocks, \emph{BeamNet} will more likely recognize the I/Q patterns in the constellation.  The input is further processed by 7 Conv layers, each followed by a maximum pooling (MaxPool) layer with filters of size 1x2, which ultimately reduce the output dimension of each Conv layer in half. Two dense layers follow the Conv + MaxPool layers, and finally a Softmax layer to obtain the probability of each beam or \gls{aoa}.

We leverage a \gls{cnn} since we want to learn short-size patterns in the I/Q constellation plane, which will ultimately distinguish different beamforming vectors \cite{restuccia2020physical}. To clearly explain this point, the top portion of Figure \ref{fig:baseline} shows the beam patterns (\textit{i.e.}, the transmitted power as a function of the azimuth angle) for beams 0, 11, and 12 of the 24-beam codebook used in our experimental testbed described in Section \ref{sec:ni_testbed}. These patterns were obtained from the testbed vendor through measurements in an anechoic chamber. Figure \ref{fig:baseline} shows that beam 0 has a very different shape with respect to beams 11 and 12. In the middle portion of Figure \ref{fig:baseline}, we also include example of different I/Q sequences for each of the beams. 

The core idea is that the CNN filters will learn to distinguish these ``imperfections'' in the I/Q constellation plane. To further verify this is the case, we have investigated how the filters in the first convolutional layer of \textit{BeamNet} react to the different beams. The bottom size of Figure \ref{fig:baseline} shows the average activation values (over the test set) for the filters in the first layer of \textit{BeamNet} that have at least one positive value (six filters in total). We notice that beams 11/12 have a strong positive reaction to filter 17, which is also very similar in magnitude. Moreover, the strongest reaction for beam 0 happens for filter 10, where beams 11/12 have a strong negative activation value. This confirms that \emph{BeamNet} is learning to distinguish beams by discriminating different patterns in the received I/Q waveform.

\section{DeepBeam Use Cases}\label{sec:applications}

As  mentioned earlier, DeepBeam is independent of the specific wireless protocol stack, since it relies on unprocessed I/Q samples and  thus can be used for any beam management operation (\textit{i.e.}, initial access, beam tracking, neighbor discovery). Nonetheless, to provide a concrete example of the effectiveness of DeepBeam, in the following paragraphs we will describe two use cases based on 5G protocol stacks, \textit{i.e.}, the initial access for 3GPP NR and neighbor discovery in mmWave vehicular networks.\smallskip

\textbf{Initial Access in 3GPP NR.}~NR is a set of specifications for 5G cellular networks defined by the 3GPP Release 15 in 2018 and Release 16 in 2020. Its physical layer is based on \gls{ofdm}, with a flexible frame structure in which the symbol duration and subcarrier spacing can be adapted to match traffic requirements. Henceforth, we consider numerology 3~\cite{38300}, with a symbol duration $T_{\rm sym} = 8.92$ $\mu$s, and slots of 14 symbols with duration $T_{\rm slot} = 250$ $\mu$s.

Beam management for the \gls{ia} procedure in 3GPP NR involves four steps~\cite{giordani2018tutorial,38300}. In the first (\textit{beam sweep}), the base station transmits directional \glspl{ss} to cover all the \glspl{txb} of a certain codebook. Notably, each beam is swept with an \gls{ssb}, which is a group of 4 \gls{ofdm} symbols and 240 subcarriers in frequency. \glspl{ssb} are interleaved to data transmissions in pre-defined time instants during bursts of 5 ms, as discussed in~\cite{moto2017ss}. There can be at most $N_{\rm SS} = 64$ \glspl{ssb} for each burst, and, if the sweep is not completed, the procedure resumes during the next burst. \gls{ss} bursts are repeated with a periodicity $T_{\rm SS}$ that can be configured by the NR protocol stack (5 to 160 ms, with default 20 ms). During the \gls{ssb} beam sweep, the \gls{ue} itself, if configured for directional reception, performs a directional scan, measuring the quality of each beam pair (second step, \textit{beam measurement}). Then, the \gls{ue} selects the beam to be used to perform initial access (third step, \textit{beam decision}). During the next \gls{ssb} in the selected direction, the \gls{ue} acquires information on the time and frequency resources in which the base station will be in receive mode for the random access message using the same \gls{txb} (fourth step, \textit{beam reporting}).

\glsreset{ebs}
Let us consider an \gls{ebs}, with $N_{tx}$ beams at the \gls{txer}, and $M_{rx}$ at the \gls{rxer}. Thus, the number of beams to be scanned is then $N_{tx} M_{rx}$. By adapting the analysis from~\cite{giordani2018tutorial} for an analog beamforming case, the time required to complete an \gls{ebs} (i.e., steps 1 and 2) with the 3GPP NR frame structure is 
\begin{equation}
  T_{\rm EBS} = T_{\rm SS}\left(\left\lceil\frac{N_{tx} M_{rx}}{N_{\rm SS}}\right\rceil - 1\right) + \hat{T}_{\rm EBS}.
  \label{eq:nr_time}
\end{equation}
The first term of the sum in Equation \ref{eq:nr_time} represents the time to scan the first $(\lceil N_{tx} M_{rx} / N_{\rm SS}\rceil - 1 )N_{\rm SS}$ \glspl{ssb}, in bursts of $N_{\rm SS}$ \glspl{ssb}. The remaining $\hat{N}_{SS} = N_{tx} M_{rx} - (\lceil N_{tx} M_{rx} / N_{\rm SS}\rceil - 1 )N_{\rm SS}$ will only occupy a portion $\hat{T}_{\rm EBS}$ of the 5 ms of the last \gls{ssb} burst, i.e., following~\cite{moto2017ss}, 
\begin{equation}
  	\hat{T}_{\rm EBS} = \begin{cases}
    \frac{\hat{N}_{SS}}{2}T_{slot} - 2T_{sym} & \mbox{ if }  \hat{N}_{SS}\bmod 2   = 0\\
    \left\lfloor\frac{\hat{N}_{SS}}{2}\right\rfloor T_{slot} + 6T_{sym} & \mbox{ otherwise.}
  \end{cases}
  \label{eq:T_last}
\end{equation} 

Thanks to the \gls{txb} and \gls{aoa} inference, as highlighted in Figure~\ref{fig:intuition}, DeepBeam can skip the \gls{ebs} by passively eavesdropping ongoing data and control transmissions between the \gls{txer} and other users. As discussed in Section~\ref{sec:learning_engine}, DeepBeam needs to acquire $\xi = K\cdot L$ I/Q samples to perform the classification task on the two inference engines. In 3GPP NR, each \gls{ofdm} symbol is composed by $S$ subcarriers, with $24 \cdot 12 \le S \le 275 \cdot 12$ subcarriers for numerology 3 (\textit{i.e.}, at most 400 MHz of bandwidth for each carrier frequency). Assuming one I/Q sample for each subcarrier (\emph{i.e.}, without considering oversampling factors), DeepBeam needs to eavesdrop $E = \lceil \xi / S \rceil$ \gls{ofdm} symbols. Eventually, considering a \gls{txer} that allocates $J$ symbols to each user in its coverage area, with a round-robin scheduler, the time required for passive data collection on the $N_{tx}$ \glspl{txb} is
\begin{equation}
	T_{\rm DB, d} = \max\{J, E\} N_{tx} T_{\rm sym}.
\end{equation}
In addition, the inference engines of DeepBeam require a certain processing time to perform the classification.\footnote{We consider the processing time negligible in the case of a traditional \gls{ebs}, as a worst-case scenario for our comparison.} The end-to-end latency of the learning engine is $T_{\rm DB, c, e2e}$, with the slowest layer providing results with a delay of $T_{\rm DB, c, max}$. When implemented on \gls{fpga}, it is possible to exploit a pipeline effect, thus the network will classify $N_{tx}$ beams in $T_{\rm DB, c, e2e} + (N_{tx} - 1) T_{\rm DB, c, max}$. Eventually, the overall delay (data collection and classification) of the DeepBeam engine for the 3GPP NR initial access is
\begin{multline}\label{eq:tdb}
 	T_{\rm DB} = \max\{\max\{J, E\} T_{\rm sym}, \, T_{\rm DB, c, max}\} (N_{tx} - 1) + \\
 	\max\{J, E\} T_{\rm sym} + T_{\rm DB, c, e2e}.
\end{multline}

\textbf{Latency Analysis with FPGA CNN Synthesis.}~To understand whether DeepBeam can truly deliver an accuracy boost with respect to existing technologies, we have synthesized in \gls{fpga} an instance of the inference engine for the \gls{txb} classification. Specifically, we have considered a \gls{cnn} with input size $\xi = 512$ I/Q samples, a single convolutional layer with 16 filters, which yields an accuracy of 90\% in a 5-beam classification problem (discussed in Section \ref{sec:codebooks}). For synthesis, we targeted a Xilinx Zynq-7000 with part number xc7z045ffg900-2, a commonly used for software-defined radio implementations \cite{openwifigithub,Fmcomms2}. We used high-level synthesis (HLS) for our \gls{cnn} design. HLS allows the conversion of a C++-level description of the \gls{cnn} directly into hardware description language (HDL) code such as Verilog. Therefore, improved results could be achieved with different design and synthesis strategies that further optimize real-time operations and minimize latency. By pipelining portions of the design, we are able to obtain $T_{\rm DB, c, e2e} = 0.492$ ms, while $T_{\rm DB, c, max} = 0.34$ ms. The resource utilization of the \gls{cnn} design is below 5\% -- specifically, our design utilizes 32/1090 block RAMs, 28/900 DSP48E, 3719/437200 flip-flops and 2875/218600 look-up tables. Resource consumption can be further brought down by avoiding pipelining, to the detriment of latency. 

\begin{figure}[t]
    \centering
    \setlength\fwidth{.8\columnwidth}
    \setlength\fheight{0.25\columnwidth}
%
%
\definecolor{mycolor1}{rgb}{0.00000,0.44700,0.74100}%
\definecolor{mycolor2}{rgb}{0.85000,0.32500,0.09800}%
\definecolor{mycolor3}{rgb}{0.92900,0.69400,0.12500}%
\definecolor{mycolor4}{rgb}{0.49400,0.18400,0.55600}%
\begin{tikzpicture}
\pgfplotsset{every tick label/.append style={font=\scriptsize}}

\begin{axis}[%
width=0.951\fwidth,
height=\fheight,
at={(0\fwidth,0\fheight)},
scale only axis,
bar shift auto,
xmin=0.5,
xmax=4.5,
xlabel style={font=\scriptsize},
ylabel style={font=\scriptsize},
xlabel={$T_{\rm SS}$ [ms]},
ymin=0,
ymax=82,
yminorticks=true,
ylabel={Beam sweep latency [ms]},
axis background/.style={fill=white},
xmajorgrids,
ymajorgrids,
yminorgrids,
xtick=data,
xticklabels={5, 10, 20, 40},
enlarge x limits=0.02,
legend style={legend cell align=left, align=left, draw=white!15!black, font=\scriptsize, at={(0.01, 0.99)}, anchor=north west},
legend columns=2,
]

\addplot [ybar, bar width=0.178, fill=mycolor2, draw=black, area legend, postaction={pattern=north east lines}]
  table[row sep=crcr]{%
1	5.7609\\
2	5.7609\\
3	5.7609\\
4	5.7609\\
};
\addlegendentry{DeepBeam, $J=1$}

\addplot [ybar, bar width=0.178, fill=mycolor3, draw=black, area legend, postaction={pattern=horizontal lines}]
  table[row sep=crcr]{%
1	5.7877\\
2	5.7877\\
3	5.7877\\
4	5.7877\\
};
\addlegendentry{DeepBeam, $J=4$}

\addplot [ybar, bar width=0.178, fill=mycolor4, draw=black, area legend]
  table[row sep=crcr]{%
1	5.8769\\
2	5.8769\\
3	5.8769\\
4	5.8769\\
};
\addlegendentry{DeepBeam, $J=14$}

\addplot [ybar, bar width=0.178, fill=mycolor1, draw=black, area legend, postaction={pattern=dots}]
  table[row sep=crcr]{%
1	10.9821625\\
2	20.9821625\\
3	40.9821625\\
4	80.9821625\\
};
\addlegendentry{3GPP NR EBS}

\coordinate (pt) at (axis cs:4.3,-3);
\coordinate (pt-low) at (axis cs:3.57,5.9);
\coordinate (pt-high) at (axis cs:4.2,5.9);

\end{axis}

\draw [larrow] (5.4,0.87) -- (pt-low);
\draw [larrow] (7.2,0.87) -- (pt-high);

\node[pin={[pin edge={mycolor1,ultra thin}]90:{%
    \begin{tikzpicture}[trim axis left,trim axis right]
    \pgfplotsset{every tick label/.append style={font=\tiny}}
    \begin{axis}[
    scale only axis,
bar shift auto,
width=0.25\fwidth,
height=0.195\fheight,
axis x line*=bottom,
axis y line*=left, 
xmin=3.6,
xmax=4.4,
xtick={4},
xticklabel style={align=center},
xticklabels={},
xmajorgrids,
ymajorgrids,
ymin=5.7,
ymax=5.91,
ytick={5.75, 5.85},
axis background/.style={fill=white},
yticklabel shift=-3pt,
    ]


\addplot [ybar, bar width=0.178, fill=mycolor2, draw=black, area legend, postaction={pattern=north east lines}]
  table[row sep=crcr]{%
4	5.7609\\
};

\addplot [ybar, bar width=0.178, fill=mycolor3, draw=black, area legend, postaction={pattern=horizontal lines}]
  table[row sep=crcr]{%
4	5.7877\\
};

\addplot [ybar, bar width=0.178, fill=mycolor4, draw=black, area legend]
  table[row sep=crcr]{%
4	5.8769\\
};

    \end{axis}
    \end{tikzpicture}%
}}] at (pt) {};

\end{tikzpicture}%
    \setlength\abovecaptionskip{0cm}
    \setlength\belowcaptionskip{-.3cm}
    \caption{Beam sweep latency, NR vs DeepBeam, for different values of the \gls{ss} burst periodicity $T_{\rm SS}$ and of the number of contiguous symbols $J$ allocated to each \gls{txb} by the base station scheduler.}
    \label{fig:ia_latency}
\end{figure}
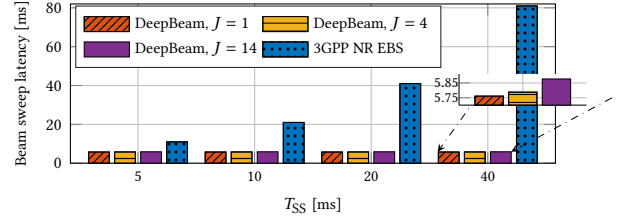

Figure~\ref{fig:ia_latency} reports $T_{\rm EBS}$ and $T_{\rm DB}$ for different values of $T_{\rm SS}$.
We assume numerology 3, a bandwidth of 400 MHz (i.e., $S = 3300$), $N_{tx}= N_{rx} = 12$, and different values of $J$, to represent various resource allocation policies of the NR base station. The results show how DeepBeam manages to decrease the beam sweep latency by a factor between $1.87$ (for $T_{\rm SS} = 5$ ms) and $14.05$ (for $T_{\rm SS} = 40$ ms). Notice that $T_{\rm SS} = 5$ ms represents a configuration where the overhead is rather high, as there is no interval between consecutive \gls{ssb} bursts. In the default configuration with $T_{\rm SS} = 20$ ms, DeepBeam reduces the latency by up to $7.11$ times. \smallskip
 
\textbf{Neighbor Discovery in Vehicular Networks}.~Beam tracking and neighbor discovery are even more challenging in vehicular scenarios~\cite{asadi2018fml}, since the dynamics of the system reduce the coherence time and prevent from efficiently using pilot signals. 
Moreover, since neighboring vehicles may change the reciprocal position frequently, each node needs fresh information on the best beam selection before starting a communication with another peer. 

\begin{figure}[b]
	\centering
	\includegraphics[width=.6\columnwidth]{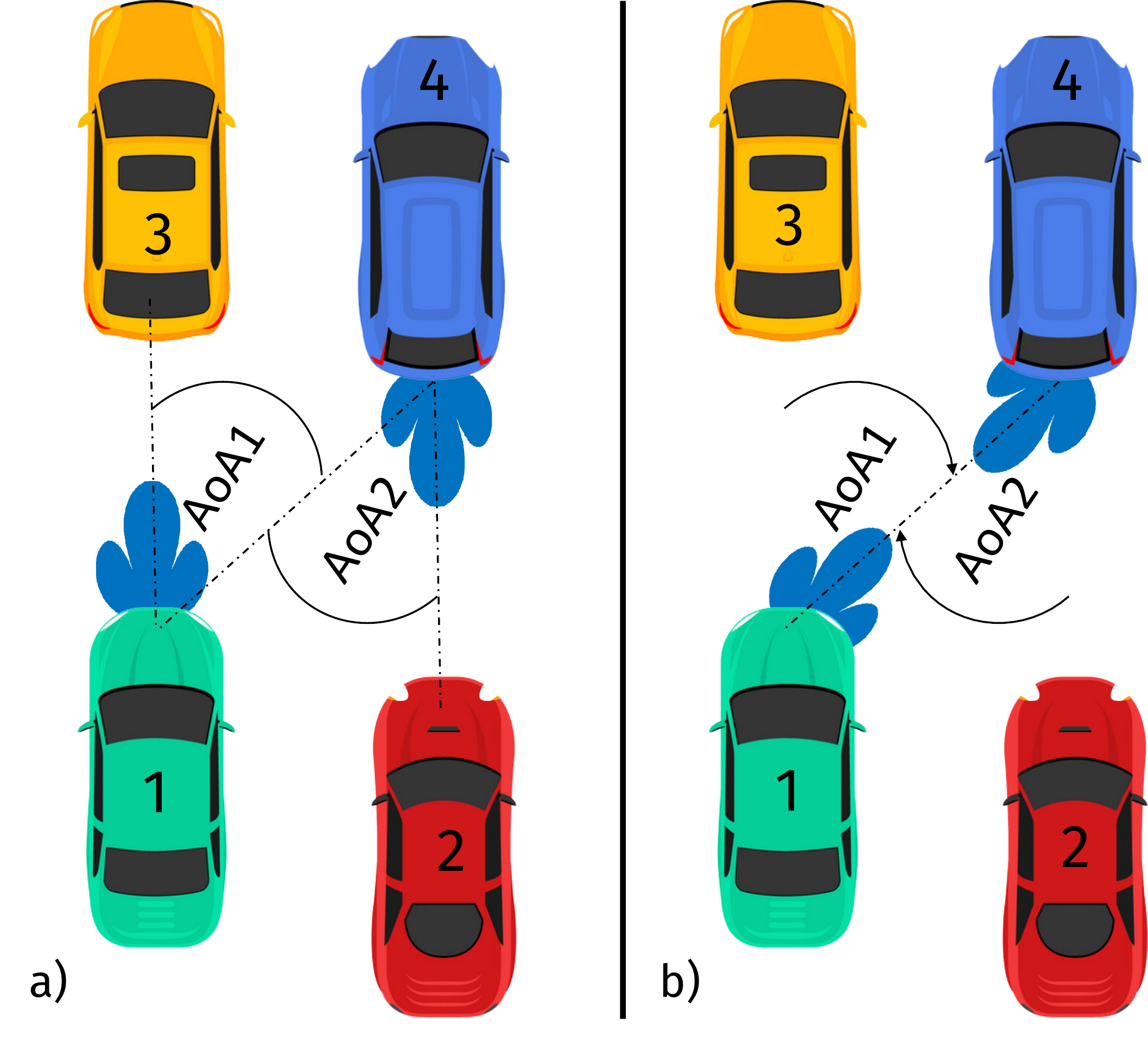}
	\caption{DeepBeam in a vehicular ad hoc scenario. (a) Vehicle 1 is communicating with vehicle 3, vehicle 2 with vehicle 4. Using DeepBeam, vehicles 1 and 4 infer the reciprocal \gls{aoa} by passively eavesdropping ongoing data transmissions. (b) Vehicles 1 and 4 can steer the beam toward each other when they need to exchange data.}
	\label{fig:adhoc}
\end{figure}

Prior work on ad hoc mmWave communications relies on contextual information, custom hardware and/or signaling to perform beam management~\cite{prelcic2016radar}. Conversely, Figure~\ref{fig:adhoc} illustrates how DeepBeam can be effective also in a \gls{mmwave} vehicular/ad hoc scenario. In this example, four vehicles are proceeding on a two-lane street, transmitting and receiving data with the vehicle in the same lane. At the same time, the vehicles can use the DeepBeam inference engine to classify the \gls{aoa} of the waveform received from the transmissions of the vehicles in the other lane. For example, if IEEE 802.11ad is used, DeepBeam can perform data collection during the interframe intervals which are mandated by the standard specifications, \textit{e.g.}, the \gls{difs} and the \gls{sifs}, which would allow the collection of 22880 and 5280 I/Q samples, respectively, over $13$ $\mu$s and $3$ $\mu$s. Moreover, as the data collection and classification can be performed while (in this example) vehicles 1 and 3 are communicating with each other, when vehicle 1 starts transmitting to vehicle 4, it is already aware of the \gls{txb} to use (\textit{i.e.}, that corresponding to the \gls{aoa} classified by DeepBeam). This makes it possible to skip any beam sweep or coordination before the link establishment between vehicle 1 and vehicle 4. Once again, if considering IEEE 802.11ad, this could take up to $225.4$ $\mu$s for a codebook with 12 beams, according to~\cite{patra2017design}.

\section{Experimental Setup and Dataset}\label{sec:experimental_setup}

This section describes the two \gls{mmwave} testbeds used to collect the waveform data (Sections \ref{sec:ni_testbed} and \ref{sec:piradio_testbed}), how our datasets are structured, and how the models were trained (Section \ref{sec:data_struct}).

\vspace{-.2cm}
\subsection{Single-RF-chain Testbed}\label{sec:ni_testbed}
\vspace{-.1cm}

\begin{table*}[t]
    \centering
    \begin{tabular}{lllll}
    \toprule
         Classification target & TX Codebook & Testbed & Configuration & (TX, RX) antenna combinations \\\midrule
         \gls{txb} & 24-beams codebook & Single-RF-chain & Basic, with obstacle, diagonal & SiBeam (0, 1), (1, 0), (2, 1), (3, 1) \\
         \gls{txb} & 12-beams codebook & Single-RF-chain & Basic, with obstacle, diagonal & SiBeam (0, 1), (1, 0), (2, 1), (3, 1) \\
         \gls{aoa} & 24-beams codebook & Single-RF-chain & Basic, with obstacle, diagonal & SiBeam (0, 1), (1, 0), (0, 2), (0, 3) \\
         \gls{txb} & 5-beams codebook & Multi-RF-chain & Multi-RF-chain basic & Node A, Node B\\
         \bottomrule
    \end{tabular}
	\setlength\abovecaptionskip{-.3cm}
    \caption{Setups for the I/Q data collection. Dataset: \url{http://hdl.handle.net/2047/D20409451}, repository: \url{https://github.com/wineslab/deepbeam}}
    \label{tab:dataset}
\end{table*}


\begin{figure}[t]
	\centering
	\includegraphics[width=.85\columnwidth]{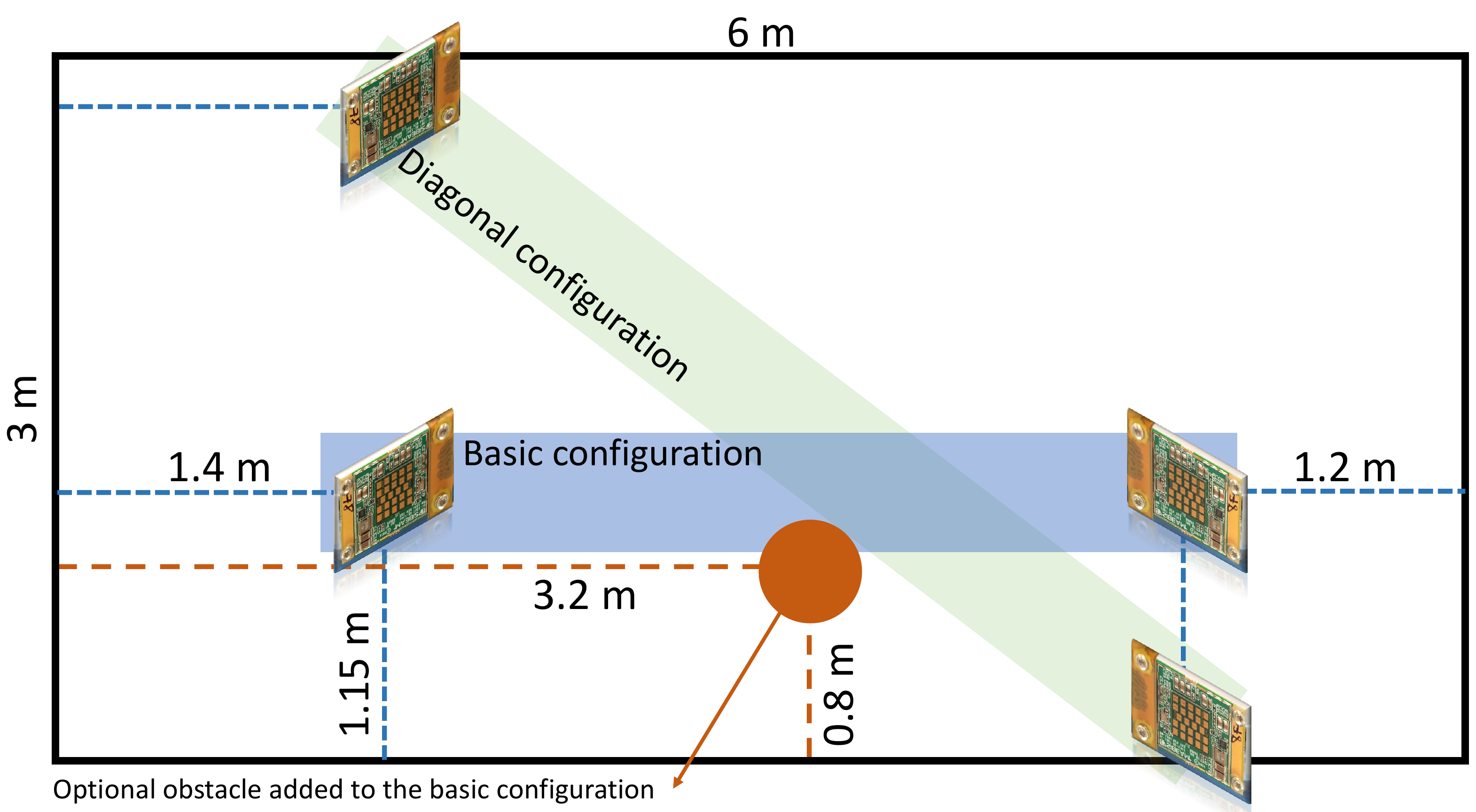}
	\setlength\abovecaptionskip{0.2cm}
	\setlength\belowcaptionskip{-.5cm}
	\caption{Configuration of the room where the single-RF-chain dataset was collected, and position of the radios in the basic and diagonal configuration, and of the obstacle.}
	\label{fig:ni}
\end{figure}

This testbed is based on the NI mmWave platform~\cite{nitestbed}, with two software-defined transceivers implemented on \glspl{fpga}, mounted on PXIe chassis, and running a custom 802.11ad-like physical layer. Besides the \glspl{fpga}, each transceiver chassis includes an \gls{adc} and a \gls{dac}, operating in baseband at 3.072 GS/s. The two nodes are equipped with 60 GHz radio frontends from SiBeam, which feature an up-conversion circuit, capable of bringing the signal to an \gls{rf} carrier of 60.48 GHz, with an \gls{rf} bandwidth of 1.76 GHz, and an analog phased array. The array (also shown in Figure~\ref{fig:ni}) has 12 antenna elements for the \gls{txer} chain, and 12 for the \gls{rxer} chain. Each element can be controlled with $4$ phase settings (i.e., a rotation of $0^\circ$, $90^\circ$, $180^\circ$, or $270^\circ$) to perform beam steering. By default, two codebooks are provided, with 24 beams in the azimuth plane, or 12 beams steered in the azimuth and elevation planes. The transmit power is 12 dBm, and it is possible to control the \gls{rxer} gain of the SiBeam boards. The physical layer in the two NI transceivers is based on IEEE 802.11ad, and generates (or receives) samples at a rate that matches that of the \gls{adc}/\gls{dac}. I/Q samples are aggregated in blocks of 2048 samples, and 150 blocks define a slot of 100 $\mu$s. 100 slots are then grouped in a frame (10 ms), which constitutes the basic transmission unit.

For the data collection, the two \gls{mmwave} nodes were positioned as in Figure~\ref{fig:ni}, in a $6 \times 3$ m room, with three different configurations. The first (\textit{i.e.}, \textit{basic configuration}) features two arrays facing each other, at a distance of $3.4$ m, and at $1.15$ m from the side wall. They have the same position in the second setup (\textit{i.e.}, \textit{obstacle configuration}), but an obstacle (i.e., a chair) is added in the space between the two antennas, without obstructing the \gls{los}. In the third setting (\textit{i.e.}, \textit{diagonal configuration}), the arrays face each other at a distance of $4.40$ m, with the link crossing the room diagonally.

\vspace{-.2cm}
\subsection{Multi-RF-chain Testbed}\label{sec:piradio_testbed}
\vspace{-.1cm}

\begin{figure}[t]
    \centering
    \includegraphics[width=.85\columnwidth]{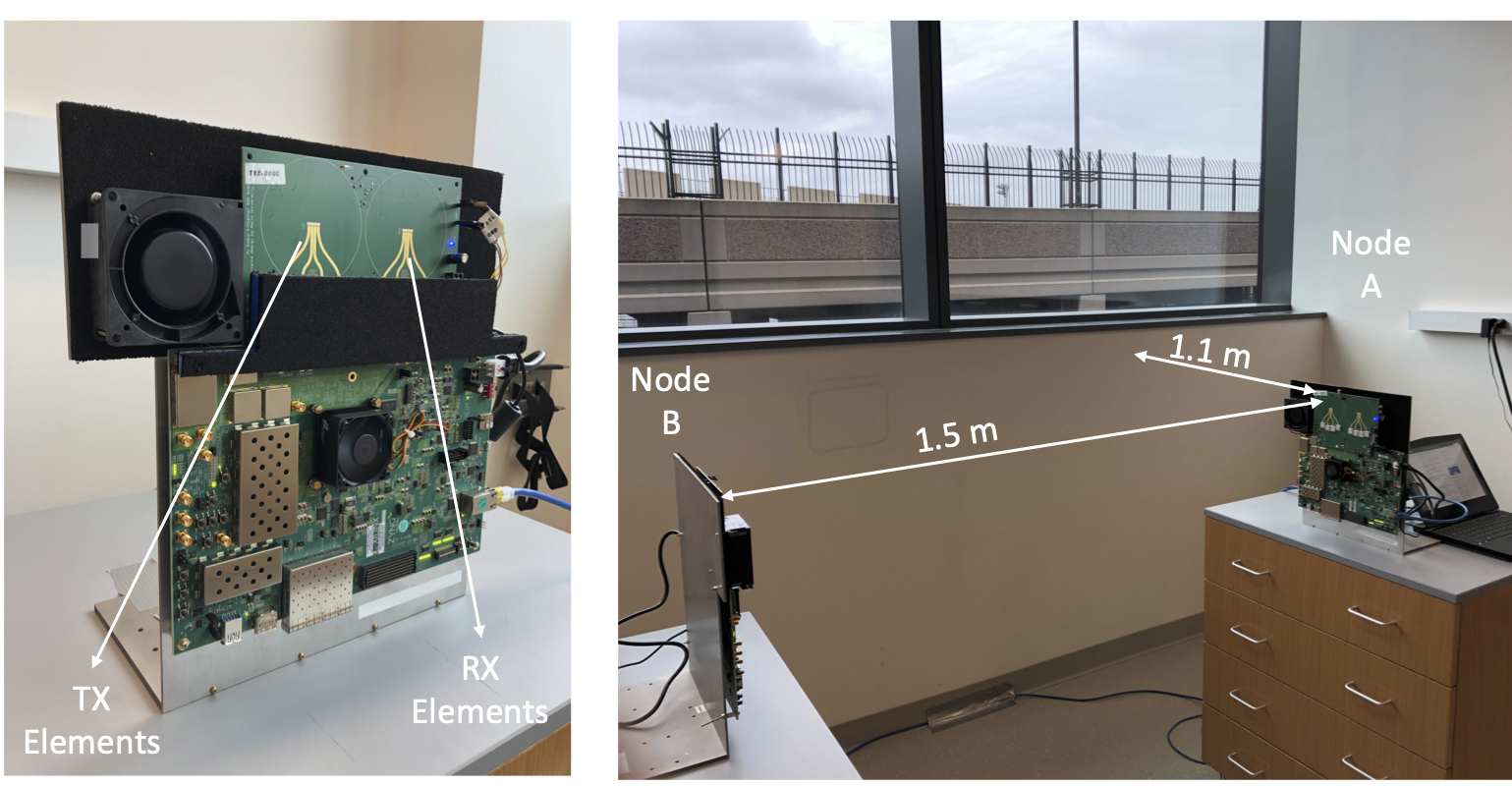}
	\setlength\abovecaptionskip{0.2cm}
    \setlength\belowcaptionskip{-.6cm}
    \caption{Multi-RF-chain testbed setup.}
    \label{fig:piradios}
\end{figure}

The second testbed features two fully-digital \gls{mmwave} transceiver boards, shown in Figure~\ref{fig:piradios}, each based on an off-the-shelf Xilinx ZCU111 RFSoC-based evaluation board and a custom mezzanine board. This takes care of the \gls{rf} up-conversion, and has two arrays (for the \gls{txer} and the \gls{rxer}) with 4 patch antenna elements each~\cite{haarla2020characterizing}. With respect to the SiBeams radios, in this setup each antenna element is connected to an \gls{rf} chain, with its own up-converters (with an output power of 12 dBm per channel), in the mezzanine board, and \glspl{adc}/\glspl{dac}, on the Xilinx RFSoC. While the sampling rate of the \glspl{adc}/\glspl{dac} is 3.93216 GS/s, with separate elements for the in-phase and quadrature components, the effective \gls{rf} bandwidth is limited to 2 GHz by the up-converters and patch antennas. We operate the boards at a carrier frequency of 58 GHz. The two transceivers use a custom physical layer, based on \gls{ofdm}, with a sampling rate that matches that of the \glspl{adc}/\gls{dac}, an oversampling factor of 4, and 256 subcarriers over a bandwidth of 1 GHz.

Differently from the single-RF-chain testbed, in which beamforming is performed in the analog domain by selecting one of the four phase shift available in each antenna elements, in the transceivers of this testbed the beamforming vector is applied digitally, \textit{i.e.}, the I/Q samples are multiplied by a vector of digital phase shifts (one for each of the 4 RF chains) before (after) the \gls{adc} (\gls{dac}) conversion. This enables the definition of more precise beam patterns, and more degrees of freedom with respect to the selection of the steering vector. The data collection for this pair of nodes was performed with the two transceivers facing each other, at a distance of $1.5$ m, as shown in Figure~\ref{fig:piradios}.

\vspace{-.2cm}
\subsection{Datasets Structure and Training Procedure}\label{sec:data_struct}
\vspace{-.1cm}

We collected more than 4 TB of raw I/Q samples to evaluate the performance of DeepBeam, using the single- and the multi-RF-chain testbeds. Table~\ref{tab:dataset} summarizes the different configurations in which the data collection was performed. Notably, for the single-RF-chain testbed, we used four different SiBeam 60 GHz frontends, the three configurations described in Figure~\ref{fig:ni}, and the two default \gls{txb} codebooks of the SiBeam phased arrays. For the \gls{aoa} dataset, we physically rotate the receive phased array by $\theta \in \{-45^\circ, 0^\circ, 45^\circ\}$ with respect to the direction between the \gls{txer} and \gls{rxer}.
To collect data with low and high \gls{snr} (\textit{i.e.,} in a range between -15 dB and 20 dB, according to the combination of \gls{txb} and gain), we consider three RX gain values for each configuration the single-RF-chain testbed, and three TX gain for the multi-RF-chain testbed. For both, the \gls{rxb} is always steered toward the boresight direction of the \gls{rxer} array. 
The raw I/Q data is collected in blocks of 2048 samples, for both the single-RF-chain and the multi-RF-chain testbed. For the first, we collected 150000 blocks for each combination of \gls{txb} and RX gain. For the second, we collected 50000 blocks for each combination of \gls{txb} and TX gain.

Our models were trained using the Adam optimizer \cite{kingma2014adam} with a learning rate of $l = 0.0001$. Our training minimizes the prediction error over the training set through back-propagation, with categorical cross-entropy as loss function. We have implemented \emph{BeamNet}, and the training/testing code in Keras, with TensorFlow as a backend. We used an NVIDA DGX system equipped with 4 Tesla V100 GPUs. We trained our models for at least ten epochs, with batch size of 100. Our dataset was split into 60\% training set and 40\% testing set. 


\section{Experimental Results}
\label{sec:results}

This section presents an extensive set of experimental results that validate the performance of DeepBeam. We first characterize the accuracy with different codebooks and different input size (Sec.~\ref{sec:codebooks}). We then present insights on DeepBeam's performance with different \gls{snr} levels and locations (Sec.~\ref{sec:snr}). The third set of results explores how DeepBeam generalizes when trained on I/Q samples from a phased array antenna, and tested on another, and investigates cross-training solutions (Sec.~\ref{sec:tota}). Finally, we discuss limitations and future extensions (Sec.~\ref{sec:discussion}).


\vspace{-.2cm}
\subsection{Accuracy Results with Different Codebooks and Input Sizes}\label{sec:codebooks}
\vspace{-.1cm}

\begin{figure}[t]
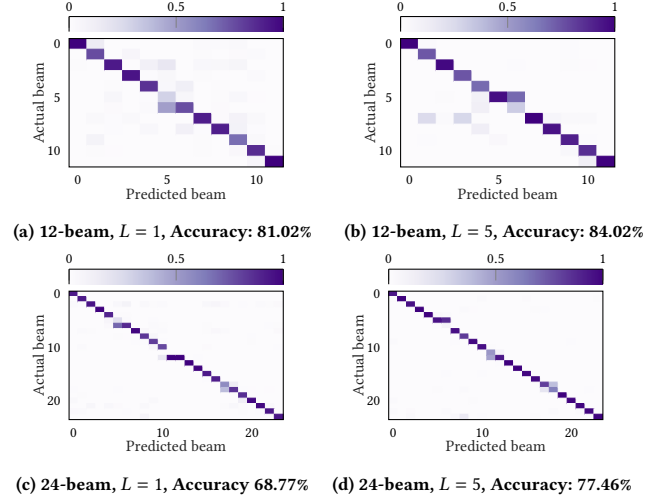

    \centering
    \begin{subfigure}[t]{0.48\columnwidth}
    \centering
    \setlength\fwidth{.7\columnwidth}
    \setlength\fheight{0.42\columnwidth}
    \begin{tikzpicture}
\pgfplotsset{every tick label/.append style={font=\tiny}}

\begin{axis}[
enlargelimits=false,
colorbar,
colormap/Purples,
width=\fwidth,
height=\fheight,
at={(0\fwidth,0\fheight)},
scale only axis,
tick align=inside,
xlabel={Predicted beam},
xmin=-0.5,
xmax=11.5,
xtick style={draw=none},
xlabel style={font=\scriptsize\color{white!15!black}},
ylabel style={font=\scriptsize\color{white!15!black}},
ylabel={Actual beam},
ymin=-0.5,
ymax=11.5,
xlabel shift=-5pt,
ylabel shift=-5pt,
ytick style={draw=none},
axis background/.style={fill=white},
colorbar horizontal,
colorbar style={
at={(0,1.05)},               
anchor=below south west,    
width=\pgfkeysvalueof{/pgfplots/parent axis width},
xtick={0, 0.5, 1},
xmin=0,
xmax=1,
axis x line*=top,
xticklabel shift=-1pt,
point meta min=0,
point meta max=1,
},
colorbar/width=2mm,
]
\addplot [matrix plot,point meta=explicit]
 coordinates {
(0,0) [0.9864166666666667] (0,1) [0.0] (0,2) [0.01075] (0,3) [0.0005833333333333334] (0,4) [0.0] (0,5) [0.0] (0,6) [0.0] (0,7) [0.00125] (0,8) [0.001] (0,9) [0.0] (0,10) [0.0] (0,11) [0.0] 

(1,0) [0.13025000000000003] (1,1) [0.7570833333333334] (1,2) [8.333333333333334e-05] (1,3) [0.0024166666666666672] (1,4) [0.0] (1,5) [0.0] (1,6) [0.0] (1,7) [0.02766666666666667] (1,8) [0.0012500000000000002] (1,9) [0.07125000000000001] (1,10) [0.010000000000000002] (1,11) [0.0] 

(2,0) [0.0] (2,1) [0.0] (2,2) [0.9085] (2,3) [0.0009166666666666666] (2,4) [0.0095] (2,5) [0.04583333333333333] (2,6) [0.013666666666666667] (2,7) [0.007333333333333333] (2,8) [0.006833333333333334] (2,9) [0.00275] (2,10) [0.004666666666666667] (2,11) [0.0] 

(3,0) [8.333333333333333e-05] (3,1) [0.048916666666666664] (3,2) [0.0] (3,3) [0.9235] (3,4) [0.0028333333333333335] (3,5) [0.0] (3,6) [0.0] (3,7) [0.024666666666666667] (3,8) [0.0] (3,9) [0.0] (3,10) [0.0] (3,11) [0.0] 

(4,0) [0.002416666666666667] (4,1) [0.0] (4,2) [0.09375] (4,3) [0.0069166666666666664] (4,4) [0.83175] (4,5) [0.00525] (4,6) [0.02033333333333333] (4,7) [0.0014166666666666668] (4,8) [0.03816666666666667] (4,9) [0.0] (4,10) [0.0] (4,11) [0.0] 

(5,0) [0.0] (5,1) [0.00025000000000000006] (5,2) [0.1604166666666667] (5,3) [0.03900000000000001] (5,4) [0.03633333333333334] (5,5) [0.2809166666666667] (5,6) [0.47925000000000006] (5,7) [0.0015833333333333335] (5,8) [0.00041666666666666675] (5,9) [0.0017500000000000003] (5,10) [8.333333333333334e-05] (5,11) [0.0] 

(6,0) [0.0] (6,1) [0.0] (6,2) [0.0005] (6,3) [8.333333333333333e-05] (6,4) [0.11] (6,5) [0.042833333333333334] (6,6) [0.751] (6,7) [0.0] (6,8) [0.09558333333333334] (6,9) [0.0] (6,10) [0.0] (6,11) [0.0] 

(7,0) [0.018500000000000003] (7,1) [0.011833333333333335] (7,2) [0.035750000000000004] (7,3) [0.008333333333333335] (7,4) [0.0008333333333333335] (7,5) [0.002166666666666667] (7,6) [0.0] (7,7) [0.8945833333333334] (7,8) [0.006250000000000001] (7,9) [0.012583333333333335] (7,10) [0.008833333333333335] (7,11) [0.0003333333333333334] 

(8,0) [0.019166666666666665] (8,1) [0.0] (8,2) [0.03866666666666667] (8,3) [0.0008333333333333334] (8,4) [0.010833333333333334] (8,5) [0.00925] (8,6) [0.014083333333333333] (8,7) [0.013833333333333333] (8,8) [0.8930833333333333] (8,9) [0.00025] (8,10) [0.0] (8,11) [0.0] 

(9,0) [0.028833333333333332] (9,1) [0.013333333333333334] (9,2) [0.056666666666666664] (9,3) [0.00375] (9,4) [0.0019166666666666666] (9,5) [0.03858333333333333] (9,6) [0.0023333333333333335] (9,7) [0.07991666666666666] (9,8) [0.08275] (9,9) [0.6589166666666667] (9,10) [0.032] (9,11) [0.001] 

(10,0) [0.00025] (10,1) [0.02025] (10,2) [0.0030833333333333333] (10,3) [0.016083333333333335] (10,4) [0.0014166666666666668] (10,5) [0.004416666666666667] (10,6) [0.00025] (10,7) [0.024916666666666667] (10,8) [0.005833333333333334] (10,9) [0.023916666666666666] (10,10) [0.8495833333333334] (10,11) [0.05] 

(11,0) [0.0] (11,1) [0.0] (11,2) [8.333333333333333e-05] (11,3) [0.0] (11,4) [8.333333333333333e-05] (11,5) [0.0] (11,6) [0.0006666666666666666] (11,7) [0.0004166666666666667] (11,8) [0.0005] (11,9) [0.0] (11,10) [0.010833333333333334] (11,11) [0.9874166666666667] 

};
\end{axis}
\end{tikzpicture}
    \caption{12-beam, $L=1$, Accuracy: 81.02\%}
    \label{fig:beam_pattern_12_1}
    \end{subfigure}
    \hfill
    \begin{subfigure}[t]{0.48\columnwidth}
    \centering
    \setlength\fwidth{.7\columnwidth}
    \setlength\fheight{0.42\columnwidth}
    \begin{tikzpicture}
\pgfplotsset{every tick label/.append style={font=\tiny}}

\begin{axis}[
enlargelimits=false,
colorbar,
colormap/Purples,
width=\fwidth,
height=\fheight,
at={(0\fwidth,0\fheight)},
scale only axis,
tick align=inside,
xlabel={Predicted beam},
xmin=-0.5,
xmax=11.5,
xtick style={draw=none},
xlabel style={font=\scriptsize\color{white!15!black}},
ylabel style={font=\scriptsize\color{white!15!black}},
ylabel={Actual beam},
ymin=-0.5,
ymax=11.5,
xlabel shift=-5pt,
ylabel shift=-5pt,
ytick style={draw=none},
axis background/.style={fill=white},
colorbar horizontal,
colorbar style={
at={(0,1.05)},               
anchor=below south west,    
width=\pgfkeysvalueof{/pgfplots/parent axis width},
xtick={0, 0.5, 1},
xmin=0,
xmax=1,
axis x line*=top,
xticklabel shift=-1pt,
point meta min=0,
point meta max=1,
},
colorbar/width=2mm,
]
\addplot [matrix plot,point meta=explicit]
 coordinates {
(0,0) [0.9988333333333334] (0,1) [0.0] (0,2) [0.0010833333333333333] (0,3) [0.0] (0,4) [0.0] (0,5) [0.0] (0,6) [0.0] (0,7) [8.333333333333333e-05] (0,8) [0.0] (0,9) [0.0] (0,10) [0.0] (0,11) [0.0] 

(1,0) [0.02025] (1,1) [0.7335833333333334] (1,2) [0.0] (1,3) [0.0] (1,4) [0.0] (1,5) [0.0] (1,6) [0.0] (1,7) [0.24066666666666667] (1,8) [0.0] (1,9) [0.005333333333333333] (1,10) [0.00016666666666666666] (1,11) [0.0] 

(2,0) [0.0] (2,1) [0.0] (2,2) [0.9744166666666665] (2,3) [0.0] (2,4) [0.0] (2,5) [0.00958333333333333] (2,6) [0.0] (2,7) [0.014499999999999997] (2,8) [0.0] (2,9) [0.0005833333333333333] (2,10) [0.0008333333333333332] (2,11) [8.333333333333332e-05] 

(3,0) [0.0] (3,1) [0.0] (3,2) [0.0] (3,3) [0.7376666666666667] (3,4) [0.0016666666666666668] (3,5) [0.0] (3,6) [0.0] (3,7) [0.26066666666666666] (3,8) [0.0] (3,9) [0.0] (3,10) [0.0] (3,11) [0.0] 

(4,0) [0.0] (4,1) [0.0] (4,2) [0.0012500000000000002] (4,3) [0.0] (4,4) [0.6775833333333334] (4,5) [0.09375000000000001] (4,6) [0.10341666666666668] (4,7) [0.009750000000000002] (4,8) [0.11425000000000002] (4,9) [0.0] (4,10) [0.0] (4,11) [0.0] 

(5,0) [0.0] (5,1) [0.0] (5,2) [0.0025833333333333333] (5,3) [0.0] (5,4) [0.0] (5,5) [0.9581666666666667] (5,6) [0.0035833333333333333] (5,7) [0.00025] (5,8) [0.0] (5,9) [0.03241666666666667] (5,10) [0.003] (5,11) [0.0] 

(6,0) [0.0] (6,1) [0.0] (6,2) [0.0065] (6,3) [0.0] (6,4) [8.333333333333333e-05] (6,5) [0.6821666666666667] (6,6) [0.30883333333333335] (6,7) [8.333333333333333e-05] (6,8) [0.0018333333333333333] (6,9) [0.0005] (6,10) [0.0] (6,11) [0.0] 

(7,0) [0.0] (7,1) [0.0] (7,2) [8.333333333333333e-05] (7,3) [0.0] (7,4) [0.0] (7,5) [0.0] (7,6) [0.0] (7,7) [0.9999166666666667] (7,8) [0.0] (7,9) [0.0] (7,10) [0.0] (7,11) [0.0] 

(8,0) [0.00016666666666666666] (8,1) [0.0] (8,2) [8.333333333333333e-05] (8,3) [0.0] (8,4) [0.0] (8,5) [8.333333333333333e-05] (8,6) [0.00025] (8,7) [0.05341666666666667] (8,8) [0.9433333333333334] (8,9) [0.0026666666666666666] (8,10) [0.0] (8,11) [0.0] 

(9,0) [0.0069166666666666664] (9,1) [0.007083333333333333] (9,2) [0.009666666666666667] (9,3) [0.0015833333333333333] (9,4) [0.0] (9,5) [0.0013333333333333333] (9,6) [0.00025] (9,7) [0.030583333333333334] (9,8) [0.00425] (9,9) [0.9] (9,10) [0.03783333333333333] (9,11) [0.0005] 

(10,0) [0.0] (10,1) [0.0005] (10,2) [0.004666666666666667] (10,3) [0.0003333333333333333] (10,4) [0.0] (10,5) [0.0004166666666666667] (10,6) [0.0] (10,7) [0.027] (10,8) [0.0] (10,9) [0.004833333333333334] (10,10) [0.8556666666666667] (10,11) [0.10658333333333334] 

(11,0) [0.0] (11,1) [0.0] (11,2) [0.0] (11,3) [0.0] (11,4) [0.0] (11,5) [0.0] (11,6) [0.0] (11,7) [0.0016666666666666668] (11,8) [0.0] (11,9) [0.0] (11,10) [0.002916666666666667] (11,11) [0.9954166666666666] 

};
\end{axis}
\end{tikzpicture}
    \caption{12-beam, $L=5$, Accuracy: 84.02\%}
    \label{fig:beam_pattern_12_5}
    \end{subfigure}
     \\
    \centering
    \begin{subfigure}[t]{0.48\columnwidth}
    \centering
    \setlength\fwidth{.7\columnwidth}
    \setlength\fheight{0.42\columnwidth}
    \input{./latex_figs/24_beam_tm_0_l_1.tex}
    \caption{24-beam, $L=1$, Accuracy 68.77\%}
    \label{fig:beam_pattern_24_1}
    \end{subfigure}
    \hfill
    \begin{subfigure}[t]{0.48\columnwidth}
    \centering
    \setlength\fwidth{.7\columnwidth}
    \setlength\fheight{0.42\columnwidth}
    \input{./latex_figs/24_beam_tm_0_l_5.tex}
    \caption{24-beam, $L=5$, Accuracy: 77.46\%}
    \label{fig:beam_pattern_24_5}
    \end{subfigure}
    \setlength\abovecaptionskip{0.2cm}
    \setlength\belowcaptionskip{-.4cm}
    \caption{Confusion matrices for 12- and 24-beam codebook, with TX antenna 0, RX antenna 1 and the basic configuration from the single-RF-chain testbed.}
    \label{fig:cm_12_24_beam}
\end{figure}

Figure \ref{fig:cm_12_24_beam} shows the confusion matrices (CMs) obtained by training \textit{BeamNet} on the 12-beam and 24-beam codebooks, for two different values of the $L$ input parameter, and $K=2048$. The accuracy reaches above 80\% and 77\% in case of the 12-beam and 24-beam codebook, respectively. Figure \ref{fig:cm_12_24_beam} indicates that \textit{BeamNet} is very accurate in predicting the beams far from the center. However, it also hints that the model gets confused when distinguishing among the central beams (11/12 and 5/6, respectively), as explained in Section \ref{sec:learning_engine}. 

\begin{figure}[t]
    \centering
    \begin{subfigure}[t]{0.48\columnwidth}
    \centering
    \setlength\fwidth{.7\columnwidth}
    \setlength\fheight{0.42\columnwidth}
\begin{tikzpicture}
\pgfplotsset{every tick label/.append style={font=\scriptsize}}

\definecolor{color0}{rgb}{0.12156862745098,0.466666666666667,0.705882352941177}
\definecolor{color1}{rgb}{1,0.498039215686275,0.0549019607843137}
\definecolor{color2}{rgb}{0.172549019607843,0.627450980392157,0.172549019607843}

\begin{axis}[
width=0.951\fwidth,
height=\fheight,
at={(0\fwidth,0\fheight)},
scale only axis,
bar shift auto,
xmin=0.5, 
xmax=4.5,
ymin=0, 
ymax=1,
axis background/.style={fill=white},
xmajorgrids,
ymajorgrids,
xminorgrids,
yminorgrids,
ylabel={Accuracy},
xlabel={Input size $K$},
xlabel style={font=\scriptsize},
ylabel style={font=\scriptsize},
legend style={legend cell align=left, align=left, draw=white!15!black, font=\scriptsize, at={(0.99, 1.4)}, anchor=north east},
xtick=data,
xticklabels={128, 256, 512, 1024},
enlarge x limits=0.02,
legend columns=3,
]

\addplot [ybar, bar width=0.22, fill=color2, draw=black, area legend, postaction={pattern=dots}]
table {%
1 0.6473
2 0.7765
3 0.9156
4 0.9578
};



\end{axis}

\end{tikzpicture}
    \caption{Accuracy vs input size $K$.}
    \label{fig:acc_piradio}
    \end{subfigure}
    \begin{subfigure}[t]{0.48\columnwidth}
    \centering
    \setlength\fwidth{.7\columnwidth}
    \setlength\fheight{0.42\columnwidth}
    \begin{tikzpicture}
\pgfplotsset{every tick label/.append style={font=\tiny}}

\begin{axis}[
enlargelimits=false,
colorbar,
colormap/Purples,
width=\fwidth,
height=\fheight,
at={(0\fwidth,0\fheight)},
scale only axis,
tick align=inside,
xlabel={Predicted Beam},
xmin=-0.5,
xmax=4.5,
xtick style={draw=none},
xlabel style={font=\scriptsize\color{white!15!black}},
ylabel style={font=\scriptsize\color{white!15!black}},
ylabel={Actual Beam},
ymin=-0.5,
ymax=4.5,
xlabel shift=-5pt,
ylabel shift=-5pt,
ytick style={draw=none},
axis background/.style={fill=white},
colorbar horizontal,
colorbar style={
at={(0,1.05)},               
anchor=below south west,    
width=\pgfkeysvalueof{/pgfplots/parent axis width},
xtick={0, 0.5, 1},
xmin=0,
xmax=1,
axis x line*=top,
xticklabel shift=-1pt,
point meta min=0,
point meta max=1,
},
colorbar/width=2mm,
]
\addplot [matrix plot,point meta=explicit]
 coordinates {
(0,0) [0.8672500000000001] (0,1) [0.012250000000000002] (0,2) [0.0] (0,3) [0.023500000000000004] (0,4) [0.09700000000000002] 

(1,0) [0.03875000000000001] (1,1) [0.8095000000000001] (1,2) [0.0] (1,3) [0.06300000000000001] (1,4) [0.08875000000000001] 

(2,0) [0.0] (2,1) [0.0] (2,2) [0.997] (2,3) [0.0] (2,4) [0.003] 

(3,0) [0.0005] (3,1) [0.00125] (3,2) [0.0] (3,3) [0.99825] (3,4) [0.0] 

(4,0) [0.07375] (4,1) [0.01575] (4,2) [0.0045] (4,3) [0.0] (4,4) [0.906] 

};
\end{axis}
\end{tikzpicture}
    \caption{Accuracy: 91.56\%}
    \label{fig:cm_piradio}
    \end{subfigure}
    \setlength\abovecaptionskip{0.2cm}
    \setlength\belowcaptionskip{-.6cm}
    \caption{Accuracy for a 5-beam codebook, using the multi-RF-chain testbed and a simplified \gls{cnn} configuration.}
    \label{fig:cm_piradios}
\end{figure}
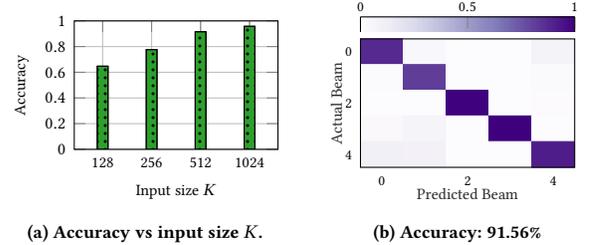

To further elaborate on the impact of the codebook and \gls{cnn} input size, we also tested DeepBeam on the dataset collected with the multi-RF-chain testbed. 
In particular,
Figure \ref{fig:acc_piradio} shows the accuracy as a function of the input size $K$. For this, we trained a smaller network than the baseline, with only one convolutional layer (with 12 filters of size 7) and no dense layer other than the softmax. Figure \ref{fig:cm_piradio} shows the \gls{cm} in the case of $K=512$, where we achieve accuracy of 91.56\%. We point out that we can achieve such high accuracy with a smaller model because in the multi-RF-chain testbed we apply beamforming vectors digitally, thus resulting in more precise beam patterns with respect to the single-RF-chain testbed. Nonetheless, this shows that DeepBeam can be applied on a heterogeneous set of devices.

\vspace{-.2cm}
\subsection{Impact of the \gls{snr} and Location}
\label{sec:snr}
\vspace{-.1cm}

\begin{figure}[t]
    \centering
    \begin{subfigure}[t]{0.48\columnwidth}
    \centering
    \setlength\fwidth{.7\columnwidth}
    \setlength\fheight{0.42\columnwidth}
    \begin{tikzpicture}
\pgfplotsset{every tick label/.append style={font=\tiny}}

\begin{axis}[
enlargelimits=false,
colorbar,
colormap/Purples,
width=\fwidth,
height=\fheight,
at={(0\fwidth,0\fheight)},
scale only axis,
tick align=inside,
xlabel={Predicted beam},
xmin=-0.5,
xmax=23.5,
xtick style={draw=none},
xlabel style={font=\scriptsize\color{white!15!black}},
ylabel style={font=\scriptsize\color{white!15!black}},
ylabel={Actual beam},
ymin=-0.5,
ymax=23.5,
xlabel shift=-5pt,
ylabel shift=-5pt,
ytick style={draw=none},
axis background/.style={fill=white},
colorbar horizontal,
colorbar style={
at={(0,1.05)},               
anchor=below south west,    
width=\pgfkeysvalueof{/pgfplots/parent axis width},
xtick={0, 0.5, 1},
xmin=0,
xmax=1,
axis x line*=top,
xticklabel shift=-1pt,
point meta min=0,
point meta max=1,
},
colorbar/width=2mm,
]
\addplot [matrix plot,point meta=explicit]
 coordinates {
(0,0) [0.469] (0,1) [0.0] (0,2) [0.0] (0,3) [0.0] (0,4) [0.0] (0,5) [0.0] (0,6) [0.0] (0,7) [0.0] (0,8) [0.16525] (0,9) [0.0] (0,10) [0.0] (0,11) [0.0] (0,12) [0.0] (0,13) [0.0] (0,14) [0.0] (0,15) [0.0] (0,16) [0.0] (0,17) [0.0] (0,18) [0.0] (0,19) [0.36575] (0,20) [0.0] (0,21) [0.0] (0,22) [0.0] (0,23) [0.0] 

(1,0) [0.017] (1,1) [0.022] (1,2) [0.0] (1,3) [0.0] (1,4) [0.0] (1,5) [0.0] (1,6) [0.0] (1,7) [0.0] (1,8) [0.87575] (1,9) [0.0] (1,10) [0.0] (1,11) [0.0] (1,12) [0.0] (1,13) [0.0] (1,14) [0.0] (1,15) [0.01375] (1,16) [0.00525] (1,17) [0.0] (1,18) [0.06625] (1,19) [0.0] (1,20) [0.0] (1,21) [0.0] (1,22) [0.0] (1,23) [0.0] 

(2,0) [0.0] (2,1) [0.0] (2,2) [0.0] (2,3) [0.97325] (2,4) [0.0] (2,5) [0.0] (2,6) [0.0] (2,7) [0.0] (2,8) [0.0] (2,9) [0.0] (2,10) [0.0] (2,11) [0.0] (2,12) [0.0] (2,13) [0.0] (2,14) [0.0] (2,15) [0.0] (2,16) [0.0] (2,17) [0.0] (2,18) [0.0] (2,19) [0.02675] (2,20) [0.0] (2,21) [0.0] (2,22) [0.0] (2,23) [0.0] 

(3,0) [0.0] (3,1) [0.0] (3,2) [0.0] (3,3) [0.75] (3,4) [0.0] (3,5) [0.0] (3,6) [0.0] (3,7) [0.0] (3,8) [0.0] (3,9) [0.0] (3,10) [0.0] (3,11) [0.0] (3,12) [0.0] (3,13) [0.0] (3,14) [0.0] (3,15) [0.0] (3,16) [0.0] (3,17) [0.0] (3,18) [0.0] (3,19) [0.25] (3,20) [0.0] (3,21) [0.0] (3,22) [0.0] (3,23) [0.0] 

(4,0) [0.0] (4,1) [0.0] (4,2) [0.0] (4,3) [0.0] (4,4) [0.0335] (4,5) [0.84775] (4,6) [0.0] (4,7) [0.00025] (4,8) [0.0] (4,9) [0.0] (4,10) [0.0] (4,11) [0.0] (4,12) [0.0] (4,13) [0.0] (4,14) [0.0] (4,15) [0.0] (4,16) [0.02275] (4,17) [0.0] (4,18) [0.0] (4,19) [0.0] (4,20) [0.0] (4,21) [0.0] (4,22) [0.0] (4,23) [0.09575] 

(5,0) [0.0] (5,1) [0.0] (5,2) [0.0] (5,3) [0.0] (5,4) [0.0015000000000000002] (5,5) [0.28025000000000005] (5,6) [0.0] (5,7) [0.0] (5,8) [0.0] (5,9) [0.0] (5,10) [0.0] (5,11) [0.0] (5,12) [0.0] (5,13) [0.0] (5,14) [0.0] (5,15) [0.0] (5,16) [0.037500000000000006] (5,17) [0.0] (5,18) [0.0] (5,19) [0.0] (5,20) [0.0] (5,21) [0.0] (5,22) [0.0] (5,23) [0.6807500000000001] 

(6,0) [0.0] (6,1) [0.0] (6,2) [0.0] (6,3) [0.0] (6,4) [0.00125] (6,5) [0.0005] (6,6) [0.5885] (6,7) [0.2255] (6,8) [0.0] (6,9) [0.0] (6,10) [0.0] (6,11) [0.0] (6,12) [0.0] (6,13) [0.0] (6,14) [0.0] (6,15) [0.0] (6,16) [0.0] (6,17) [0.0] (6,18) [0.0] (6,19) [0.0] (6,20) [0.18425] (6,21) [0.0] (6,22) [0.0] (6,23) [0.0] 

(7,0) [0.0] (7,1) [0.0] (7,2) [0.0] (7,3) [0.0] (7,4) [0.002] (7,5) [0.947] (7,6) [0.0] (7,7) [0.01825] (7,8) [0.0] (7,9) [0.0] (7,10) [0.0] (7,11) [0.0] (7,12) [0.0] (7,13) [0.0] (7,14) [0.0] (7,15) [0.0] (7,16) [0.00075] (7,17) [0.0] (7,18) [0.0] (7,19) [0.0] (7,20) [0.0] (7,21) [0.0] (7,22) [0.0] (7,23) [0.032] 

(8,0) [0.6040000000000001] (8,1) [0.0] (8,2) [0.0] (8,3) [0.0] (8,4) [0.0] (8,5) [0.0] (8,6) [0.0] (8,7) [0.0] (8,8) [0.23225000000000004] (8,9) [0.0] (8,10) [0.0] (8,11) [0.0] (8,12) [0.0] (8,13) [0.0] (8,14) [0.0] (8,15) [0.0] (8,16) [0.0] (8,17) [0.0] (8,18) [0.0] (8,19) [0.16250000000000003] (8,20) [0.0] (8,21) [0.0012500000000000002] (8,22) [0.0] (8,23) [0.0] 

(9,0) [0.0] (9,1) [0.0] (9,2) [0.0] (9,3) [0.00075] (9,4) [0.0] (9,5) [0.0] (9,6) [0.00275] (9,7) [0.0] (9,8) [0.0] (9,9) [0.9805] (9,10) [0.0] (9,11) [0.0] (9,12) [0.0] (9,13) [0.0] (9,14) [0.0] (9,15) [0.0] (9,16) [0.0] (9,17) [0.00075] (9,18) [0.0] (9,19) [0.0] (9,20) [0.01525] (9,21) [0.0] (9,22) [0.0] (9,23) [0.0] 

(10,0) [0.0] (10,1) [0.031000000000000003] (10,2) [0.0] (10,3) [0.0] (10,4) [0.0] (10,5) [0.0] (10,6) [0.0] (10,7) [0.0] (10,8) [0.04125000000000001] (10,9) [0.0] (10,10) [0.27375000000000005] (10,11) [0.00025000000000000006] (10,12) [0.0] (10,13) [0.0] (10,14) [0.008500000000000002] (10,15) [0.005000000000000001] (10,16) [0.48250000000000004] (10,17) [0.0] (10,18) [0.14325000000000002] (10,19) [0.0010000000000000002] (10,20) [0.0] (10,21) [0.0] (10,22) [0.0] (10,23) [0.013500000000000002] 

(11,0) [0.0] (11,1) [0.0] (11,2) [0.0] (11,3) [0.0] (11,4) [0.0] (11,5) [0.0] (11,6) [0.0] (11,7) [0.0] (11,8) [0.0] (11,9) [0.0] (11,10) [0.0] (11,11) [0.96975] (11,12) [0.021] (11,13) [0.00925] (11,14) [0.0] (11,15) [0.0] (11,16) [0.0] (11,17) [0.0] (11,18) [0.0] (11,19) [0.0] (11,20) [0.0] (11,21) [0.0] (11,22) [0.0] (11,23) [0.0] 

(12,0) [0.0] (12,1) [0.0] (12,2) [0.0] (12,3) [0.0] (12,4) [0.0] (12,5) [0.0] (12,6) [0.0] (12,7) [0.0] (12,8) [0.0] (12,9) [0.0005] (12,10) [0.0] (12,11) [0.96425] (12,12) [0.02925] (12,13) [0.006] (12,14) [0.0] (12,15) [0.0] (12,16) [0.0] (12,17) [0.0] (12,18) [0.0] (12,19) [0.0] (12,20) [0.0] (12,21) [0.0] (12,22) [0.0] (12,23) [0.0] 

(13,0) [0.0] (13,1) [0.0] (13,2) [0.0] (13,3) [0.0] (13,4) [0.0] (13,5) [0.0] (13,6) [0.0] (13,7) [0.0] (13,8) [0.0] (13,9) [0.0] (13,10) [0.0] (13,11) [0.12775] (13,12) [0.0] (13,13) [0.87225] (13,14) [0.0] (13,15) [0.0] (13,16) [0.0] (13,17) [0.0] (13,18) [0.0] (13,19) [0.0] (13,20) [0.0] (13,21) [0.0] (13,22) [0.0] (13,23) [0.0] 

(14,0) [0.0] (14,1) [0.0] (14,2) [0.0] (14,3) [0.0] (14,4) [0.0] (14,5) [0.00025] (14,6) [0.0] (14,7) [0.0] (14,8) [0.0] (14,9) [0.0005] (14,10) [0.0] (14,11) [0.14725] (14,12) [0.00125] (14,13) [0.8435] (14,14) [0.0055] (14,15) [0.0] (14,16) [0.0] (14,17) [0.0] (14,18) [0.0] (14,19) [0.0] (14,20) [0.001] (14,21) [0.0] (14,22) [0.0] (14,23) [0.00075] 

(15,0) [0.00025] (15,1) [0.0] (15,2) [0.0] (15,3) [0.0] (15,4) [0.0] (15,5) [0.0] (15,6) [0.0] (15,7) [0.0] (15,8) [0.0] (15,9) [0.0] (15,10) [0.0] (15,11) [0.0] (15,12) [0.0] (15,13) [0.0] (15,14) [0.0] (15,15) [0.98025] (15,16) [0.0] (15,17) [0.0] (15,18) [0.0195] (15,19) [0.0] (15,20) [0.0] (15,21) [0.0] (15,22) [0.0] (15,23) [0.0] 

(16,0) [0.0] (16,1) [0.0025000000000000005] (16,2) [0.0] (16,3) [0.0] (16,4) [0.0] (16,5) [0.0] (16,6) [0.0] (16,7) [0.0] (16,8) [0.0055000000000000005] (16,9) [0.0] (16,10) [0.0] (16,11) [0.0] (16,12) [0.0] (16,13) [0.0] (16,14) [0.0] (16,15) [0.0] (16,16) [0.8887500000000002] (16,17) [0.0] (16,18) [0.029500000000000002] (16,19) [0.0] (16,20) [0.0] (16,21) [0.0] (16,22) [0.0] (16,23) [0.07375000000000001] 

(17,0) [0.0] (17,1) [0.0] (17,2) [0.0] (17,3) [0.99075] (17,4) [0.0] (17,5) [0.0] (17,6) [0.0] (17,7) [0.0] (17,8) [0.0] (17,9) [0.0] (17,10) [0.0] (17,11) [0.0] (17,12) [0.0] (17,13) [0.0] (17,14) [0.0] (17,15) [0.0] (17,16) [0.0] (17,17) [0.00925] (17,18) [0.0] (17,19) [0.0] (17,20) [0.0] (17,21) [0.0] (17,22) [0.0] (17,23) [0.0] 

(18,0) [0.0] (18,1) [0.0] (18,2) [0.0] (18,3) [0.0] (18,4) [0.0] (18,5) [0.0] (18,6) [0.0] (18,7) [0.0] (18,8) [0.0] (18,9) [0.0] (18,10) [0.0] (18,11) [0.0] (18,12) [0.0] (18,13) [0.0] (18,14) [0.0] (18,15) [0.0] (18,16) [0.0] (18,17) [0.0] (18,18) [1.0] (18,19) [0.0] (18,20) [0.0] (18,21) [0.0] (18,22) [0.0] (18,23) [0.0] 

(19,0) [0.003] (19,1) [0.0] (19,2) [0.0] (19,3) [0.0015] (19,4) [0.0] (19,5) [0.0] (19,6) [0.0] (19,7) [0.0] (19,8) [0.0] (19,9) [0.0] (19,10) [0.0] (19,11) [0.0] (19,12) [0.0] (19,13) [0.0] (19,14) [0.0] (19,15) [0.0] (19,16) [0.0] (19,17) [0.0] (19,18) [0.0] (19,19) [0.9955] (19,20) [0.0] (19,21) [0.0] (19,22) [0.0] (19,23) [0.0] 

(20,0) [0.0] (20,1) [0.0] (20,2) [0.0] (20,3) [0.0] (20,4) [0.0012500000000000002] (20,5) [0.021750000000000002] (20,6) [0.15775000000000003] (20,7) [0.6645000000000001] (20,8) [0.0] (20,9) [0.0] (20,10) [0.0] (20,11) [0.0] (20,12) [0.0] (20,13) [0.0] (20,14) [0.0] (20,15) [0.0] (20,16) [0.0] (20,17) [0.0] (20,18) [0.0] (20,19) [0.0] (20,20) [0.15475000000000003] (20,21) [0.0] (20,22) [0.0] (20,23) [0.0] 

(21,0) [0.00025] (21,1) [0.0] (21,2) [0.0] (21,3) [0.007] (21,4) [0.0] (21,5) [0.0] (21,6) [0.0] (21,7) [0.0] (21,8) [0.0] (21,9) [0.0] (21,10) [0.0] (21,11) [0.0] (21,12) [0.0] (21,13) [0.0] (21,14) [0.0] (21,15) [0.0] (21,16) [0.0] (21,17) [0.0] (21,18) [0.0] (21,19) [0.99275] (21,20) [0.0] (21,21) [0.0] (21,22) [0.0] (21,23) [0.0] 

(22,0) [0.0] (22,1) [0.0] (22,2) [0.0] (22,3) [0.0] (22,4) [0.001] (22,5) [0.968] (22,6) [0.0] (22,7) [0.00175] (22,8) [0.0] (22,9) [0.0] (22,10) [0.0] (22,11) [0.0] (22,12) [0.0] (22,13) [0.0] (22,14) [0.0] (22,15) [0.0] (22,16) [0.00125] (22,17) [0.0] (22,18) [0.0] (22,19) [0.0] (22,20) [0.0] (22,21) [0.0] (22,22) [0.0] (22,23) [0.028] 

(23,0) [0.0] (23,1) [0.0] (23,2) [0.0] (23,3) [0.0] (23,4) [0.0] (23,5) [0.0] (23,6) [0.0] (23,7) [0.0] (23,8) [0.0] (23,9) [0.0] (23,10) [0.0] (23,11) [0.0] (23,12) [0.0] (23,13) [0.0] (23,14) [0.0] (23,15) [0.0] (23,16) [0.1145] (23,17) [0.0] (23,18) [0.00525] (23,19) [0.0] (23,20) [0.0] (23,21) [0.0] (23,22) [0.0] (23,23) [0.88025] 

};
\end{axis}
\end{tikzpicture}
    \caption{Low SNR. Accuracy 43.47\%}
    \label{fig:beam_pattern_l_snr}
    \end{subfigure}
    \hfill
    \begin{subfigure}[t]{0.48\columnwidth}
    \centering
    \setlength\fwidth{.7\columnwidth}
    \setlength\fheight{0.42\columnwidth}
    \begin{tikzpicture}
\pgfplotsset{every tick label/.append style={font=\tiny}}

\begin{axis}[
enlargelimits=false,
colorbar,
colormap/Purples,
width=\fwidth,
height=\fheight,
at={(0\fwidth,0\fheight)},
scale only axis,
tick align=inside,
xlabel={Predicted beam},
xmin=-0.5,
xmax=23.5,
xtick style={draw=none},
xlabel style={font=\scriptsize\color{white!15!black}},
ylabel style={font=\scriptsize\color{white!15!black}},
ylabel={Actual beam},
ymin=-0.5,
ymax=23.5,
xlabel shift=-5pt,
ylabel shift=-5pt,
ytick style={draw=none},
axis background/.style={fill=white},
colorbar horizontal,
colorbar style={
at={(0,1.05)},               
anchor=below south west,    
width=\pgfkeysvalueof{/pgfplots/parent axis width},
xtick={0, 0.5, 1},
xmin=0,
xmax=1,
axis x line*=top,
xticklabel shift=-1pt,
point meta min=0,
point meta max=1,
},
colorbar/width=2mm,
]
\addplot [matrix plot,point meta=explicit]
 coordinates {
(0,0) [0.9995] (0,1) [0.0] (0,2) [0.0] (0,3) [0.0] (0,4) [0.0] (0,5) [0.0] (0,6) [0.0] (0,7) [0.0] (0,8) [0.0] (0,9) [0.0] (0,10) [0.0] (0,11) [0.0] (0,12) [0.0] (0,13) [0.0] (0,14) [0.0] (0,15) [0.0] (0,16) [0.0] (0,17) [0.0] (0,18) [0.0] (0,19) [0.0005] (0,20) [0.0] (0,21) [0.0] (0,22) [0.0] (0,23) [0.0] 

(1,0) [0.0] (1,1) [1.0] (1,2) [0.0] (1,3) [0.0] (1,4) [0.0] (1,5) [0.0] (1,6) [0.0] (1,7) [0.0] (1,8) [0.0] (1,9) [0.0] (1,10) [0.0] (1,11) [0.0] (1,12) [0.0] (1,13) [0.0] (1,14) [0.0] (1,15) [0.0] (1,16) [0.0] (1,17) [0.0] (1,18) [0.0] (1,19) [0.0] (1,20) [0.0] (1,21) [0.0] (1,22) [0.0] (1,23) [0.0] 

(2,0) [0.0] (2,1) [0.0] (2,2) [1.0] (2,3) [0.0] (2,4) [0.0] (2,5) [0.0] (2,6) [0.0] (2,7) [0.0] (2,8) [0.0] (2,9) [0.0] (2,10) [0.0] (2,11) [0.0] (2,12) [0.0] (2,13) [0.0] (2,14) [0.0] (2,15) [0.0] (2,16) [0.0] (2,17) [0.0] (2,18) [0.0] (2,19) [0.0] (2,20) [0.0] (2,21) [0.0] (2,22) [0.0] (2,23) [0.0] 

(3,0) [0.0] (3,1) [0.0] (3,2) [0.0] (3,3) [0.7285] (3,4) [0.0] (3,5) [0.0] (3,6) [0.0] (3,7) [0.0] (3,8) [0.0] (3,9) [0.0] (3,10) [0.0] (3,11) [0.0] (3,12) [0.0] (3,13) [0.0] (3,14) [0.0] (3,15) [0.0] (3,16) [0.0] (3,17) [0.2715] (3,18) [0.0] (3,19) [0.0] (3,20) [0.0] (3,21) [0.0] (3,22) [0.0] (3,23) [0.0] 

(4,0) [0.0] (4,1) [0.0] (4,2) [0.0] (4,3) [0.0] (4,4) [1.0] (4,5) [0.0] (4,6) [0.0] (4,7) [0.0] (4,8) [0.0] (4,9) [0.0] (4,10) [0.0] (4,11) [0.0] (4,12) [0.0] (4,13) [0.0] (4,14) [0.0] (4,15) [0.0] (4,16) [0.0] (4,17) [0.0] (4,18) [0.0] (4,19) [0.0] (4,20) [0.0] (4,21) [0.0] (4,22) [0.0] (4,23) [0.0] 

(5,0) [0.0] (5,1) [0.0] (5,2) [0.0] (5,3) [0.0] (5,4) [0.0] (5,5) [0.006] (5,6) [0.994] (5,7) [0.0] (5,8) [0.0] (5,9) [0.0] (5,10) [0.0] (5,11) [0.0] (5,12) [0.0] (5,13) [0.0] (5,14) [0.0] (5,15) [0.0] (5,16) [0.0] (5,17) [0.0] (5,18) [0.0] (5,19) [0.0] (5,20) [0.0] (5,21) [0.0] (5,22) [0.0] (5,23) [0.0] 

(6,0) [0.0] (6,1) [0.0] (6,2) [0.0] (6,3) [0.0] (6,4) [0.0] (6,5) [0.00225] (6,6) [0.99775] (6,7) [0.0] (6,8) [0.0] (6,9) [0.0] (6,10) [0.0] (6,11) [0.0] (6,12) [0.0] (6,13) [0.0] (6,14) [0.0] (6,15) [0.0] (6,16) [0.0] (6,17) [0.0] (6,18) [0.0] (6,19) [0.0] (6,20) [0.0] (6,21) [0.0] (6,22) [0.0] (6,23) [0.0] 

(7,0) [0.0] (7,1) [0.0] (7,2) [0.0] (7,3) [0.0] (7,4) [0.0] (7,5) [0.0] (7,6) [0.0] (7,7) [1.0] (7,8) [0.0] (7,9) [0.0] (7,10) [0.0] (7,11) [0.0] (7,12) [0.0] (7,13) [0.0] (7,14) [0.0] (7,15) [0.0] (7,16) [0.0] (7,17) [0.0] (7,18) [0.0] (7,19) [0.0] (7,20) [0.0] (7,21) [0.0] (7,22) [0.0] (7,23) [0.0] 

(8,0) [0.0] (8,1) [0.0] (8,2) [0.0] (8,3) [0.0] (8,4) [0.0] (8,5) [0.0] (8,6) [0.0] (8,7) [0.0] (8,8) [1.0] (8,9) [0.0] (8,10) [0.0] (8,11) [0.0] (8,12) [0.0] (8,13) [0.0] (8,14) [0.0] (8,15) [0.0] (8,16) [0.0] (8,17) [0.0] (8,18) [0.0] (8,19) [0.0] (8,20) [0.0] (8,21) [0.0] (8,22) [0.0] (8,23) [0.0] 

(9,0) [0.0] (9,1) [0.0] (9,2) [0.0] (9,3) [0.0] (9,4) [0.0] (9,5) [0.0] (9,6) [0.0] (9,7) [0.0] (9,8) [0.0] (9,9) [1.0] (9,10) [0.0] (9,11) [0.0] (9,12) [0.0] (9,13) [0.0] (9,14) [0.0] (9,15) [0.0] (9,16) [0.0] (9,17) [0.0] (9,18) [0.0] (9,19) [0.0] (9,20) [0.0] (9,21) [0.0] (9,22) [0.0] (9,23) [0.0] 

(10,0) [0.0] (10,1) [0.0] (10,2) [0.0] (10,3) [0.0] (10,4) [0.0] (10,5) [0.0] (10,6) [0.0] (10,7) [0.0] (10,8) [0.0] (10,9) [0.0] (10,10) [0.9995] (10,11) [0.0] (10,12) [0.0] (10,13) [0.0] (10,14) [0.0005] (10,15) [0.0] (10,16) [0.0] (10,17) [0.0] (10,18) [0.0] (10,19) [0.0] (10,20) [0.0] (10,21) [0.0] (10,22) [0.0] (10,23) [0.0] 

(11,0) [0.0] (11,1) [0.0] (11,2) [0.0] (11,3) [0.0] (11,4) [0.0] (11,5) [0.0] (11,6) [0.0] (11,7) [0.0] (11,8) [0.0] (11,9) [0.0] (11,10) [0.00075] (11,11) [0.9975] (11,12) [0.0] (11,13) [0.0] (11,14) [0.00175] (11,15) [0.0] (11,16) [0.0] (11,17) [0.0] (11,18) [0.0] (11,19) [0.0] (11,20) [0.0] (11,21) [0.0] (11,22) [0.0] (11,23) [0.0] 

(12,0) [0.0] (12,1) [0.0] (12,2) [0.0] (12,3) [0.0] (12,4) [0.0] (12,5) [0.0] (12,6) [0.0] (12,7) [0.0] (12,8) [0.0] (12,9) [0.0] (12,10) [0.0005000000000000001] (12,11) [0.9952500000000001] (12,12) [0.0] (12,13) [0.0] (12,14) [0.004250000000000001] (12,15) [0.0] (12,16) [0.0] (12,17) [0.0] (12,18) [0.0] (12,19) [0.0] (12,20) [0.0] (12,21) [0.0] (12,22) [0.0] (12,23) [0.0] 

(13,0) [0.0] (13,1) [0.0] (13,2) [0.0] (13,3) [0.0] (13,4) [0.0] (13,5) [0.0] (13,6) [0.0] (13,7) [0.0] (13,8) [0.0] (13,9) [0.0] (13,10) [0.0] (13,11) [0.0] (13,12) [0.0] (13,13) [1.0] (13,14) [0.0] (13,15) [0.0] (13,16) [0.0] (13,17) [0.0] (13,18) [0.0] (13,19) [0.0] (13,20) [0.0] (13,21) [0.0] (13,22) [0.0] (13,23) [0.0] 

(14,0) [0.0] (14,1) [0.0] (14,2) [0.0] (14,3) [0.0] (14,4) [0.0] (14,5) [0.0] (14,6) [0.0] (14,7) [0.0] (14,8) [0.0] (14,9) [0.0] (14,10) [0.0] (14,11) [0.0] (14,12) [0.0] (14,13) [0.0] (14,14) [1.0] (14,15) [0.0] (14,16) [0.0] (14,17) [0.0] (14,18) [0.0] (14,19) [0.0] (14,20) [0.0] (14,21) [0.0] (14,22) [0.0] (14,23) [0.0] 

(15,0) [0.0] (15,1) [0.0] (15,2) [0.0] (15,3) [0.0] (15,4) [0.0] (15,5) [0.0] (15,6) [0.0] (15,7) [0.0] (15,8) [0.0] (15,9) [0.0] (15,10) [0.0] (15,11) [0.0] (15,12) [0.0] (15,13) [0.0] (15,14) [0.0] (15,15) [1.0] (15,16) [0.0] (15,17) [0.0] (15,18) [0.0] (15,19) [0.0] (15,20) [0.0] (15,21) [0.0] (15,22) [0.0] (15,23) [0.0] 

(16,0) [0.0] (16,1) [0.0] (16,2) [0.0] (16,3) [0.0] (16,4) [0.0] (16,5) [0.0] (16,6) [0.0] (16,7) [0.0] (16,8) [0.0] (16,9) [0.0] (16,10) [0.0] (16,11) [0.0] (16,12) [0.0] (16,13) [0.0] (16,14) [0.0] (16,15) [0.0] (16,16) [1.0] (16,17) [0.0] (16,18) [0.0] (16,19) [0.0] (16,20) [0.0] (16,21) [0.0] (16,22) [0.0] (16,23) [0.0] 

(17,0) [0.0] (17,1) [0.0] (17,2) [0.0] (17,3) [0.0] (17,4) [0.0] (17,5) [0.0] (17,6) [0.0] (17,7) [0.0] (17,8) [0.0] (17,9) [0.0] (17,10) [0.0] (17,11) [0.0] (17,12) [0.0] (17,13) [0.0] (17,14) [0.0] (17,15) [0.0] (17,16) [0.0] (17,17) [1.0] (17,18) [0.0] (17,19) [0.0] (17,20) [0.0] (17,21) [0.0] (17,22) [0.0] (17,23) [0.0] 

(18,0) [0.0] (18,1) [0.0] (18,2) [0.0] (18,3) [0.0] (18,4) [0.0] (18,5) [0.0] (18,6) [0.0] (18,7) [0.0] (18,8) [0.0] (18,9) [0.0] (18,10) [0.0] (18,11) [0.0] (18,12) [0.0] (18,13) [0.0] (18,14) [0.0] (18,15) [0.0] (18,16) [0.0] (18,17) [1.0] (18,18) [0.0] (18,19) [0.0] (18,20) [0.0] (18,21) [0.0] (18,22) [0.0] (18,23) [0.0] 

(19,0) [0.0] (19,1) [0.0] (19,2) [0.0] (19,3) [0.0] (19,4) [0.0] (19,5) [0.0] (19,6) [0.0] (19,7) [0.0] (19,8) [0.0] (19,9) [0.0] (19,10) [0.0] (19,11) [0.0] (19,12) [0.0] (19,13) [0.0] (19,14) [0.0] (19,15) [0.0] (19,16) [0.0] (19,17) [0.0] (19,18) [0.0] (19,19) [1.0] (19,20) [0.0] (19,21) [0.0] (19,22) [0.0] (19,23) [0.0] 

(20,0) [0.0] (20,1) [0.0] (20,2) [0.0] (20,3) [0.0] (20,4) [0.0] (20,5) [0.0] (20,6) [0.0] (20,7) [0.0] (20,8) [0.0] (20,9) [0.0] (20,10) [0.0] (20,11) [0.0] (20,12) [0.0] (20,13) [0.0] (20,14) [0.0] (20,15) [0.0] (20,16) [0.0] (20,17) [0.0] (20,18) [0.0] (20,19) [0.0] (20,20) [1.0] (20,21) [0.0] (20,22) [0.0] (20,23) [0.0] 

(21,0) [0.0] (21,1) [0.0] (21,2) [0.0] (21,3) [0.0] (21,4) [0.0] (21,5) [0.0] (21,6) [0.0] (21,7) [0.0] (21,8) [0.0] (21,9) [0.0] (21,10) [0.0] (21,11) [0.0] (21,12) [0.0] (21,13) [0.0] (21,14) [0.0] (21,15) [0.0] (21,16) [0.0] (21,17) [0.0] (21,18) [0.0] (21,19) [0.0] (21,20) [0.0] (21,21) [1.0] (21,22) [0.0] (21,23) [0.0] 

(22,0) [0.0] (22,1) [0.0] (22,2) [0.0] (22,3) [0.0] (22,4) [0.0] (22,5) [0.0] (22,6) [0.0] (22,7) [0.0] (22,8) [0.0] (22,9) [0.0] (22,10) [0.0] (22,11) [0.0] (22,12) [0.0] (22,13) [0.0] (22,14) [0.0] (22,15) [0.0] (22,16) [0.0] (22,17) [0.0] (22,18) [0.0] (22,19) [0.0] (22,20) [0.0] (22,21) [0.0] (22,22) [1.0] (22,23) [0.0] 

(23,0) [0.0] (23,1) [0.0] (23,2) [0.0] (23,3) [0.0] (23,4) [0.0] (23,5) [0.0] (23,6) [0.0] (23,7) [0.0] (23,8) [0.0] (23,9) [0.0] (23,10) [0.0] (23,11) [0.0] (23,12) [0.0] (23,13) [0.0] (23,14) [0.0] (23,15) [0.0] (23,16) [0.0] (23,17) [0.0] (23,18) [0.0] (23,19) [0.0] (23,20) [0.0] (23,21) [0.0] (23,22) [0.0] (23,23) [1.0] 

};
\end{axis}
\end{tikzpicture}
    \caption{High SNR. Accuracy: 86.36\%}
    \label{fig:beam_pattern_h_snr}
    \end{subfigure}
    \setlength\abovecaptionskip{0.2cm}
    \setlength\belowcaptionskip{-.3cm}
    \caption{Low and High SNR, 24-beam codebook, $L=1$, with TX antenna 0, RX antenna 1 and the basic configuration from the single-RF-chain testbed.}
        \label{fig:snr}
\end{figure}
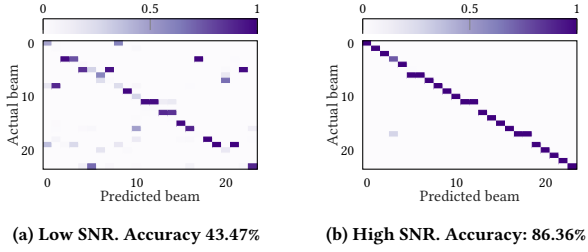

The CMs of Figure~\ref{fig:cm_12_24_beam} were obtained by mixing low, medium and high \gls{snr} waveforms. To get an insight on how the \gls{snr} impacts the accuracy of \emph{BeamNet}, Figure \ref{fig:snr} shows the CMs when low (\textit{i.e.}, below 0 dB) and high (\textit{i.e.}, above 10 dB) \gls{snr} waveforms are used to train and test the model. We only show the results for the 24-beam codebook due to space limitations. As experienced in much of existing work, Figure \ref{fig:snr} definitely indicates that there is a strong correlation between the accuracy of the model and the \gls{snr} level of the received waveforms. The accuracy drops to 43\% when low \gls{snr} samples are used, yet it goes up to 86\% when \emph{BeamNet} is trained with high \gls{snr} samples only.

\begin{figure}[t]
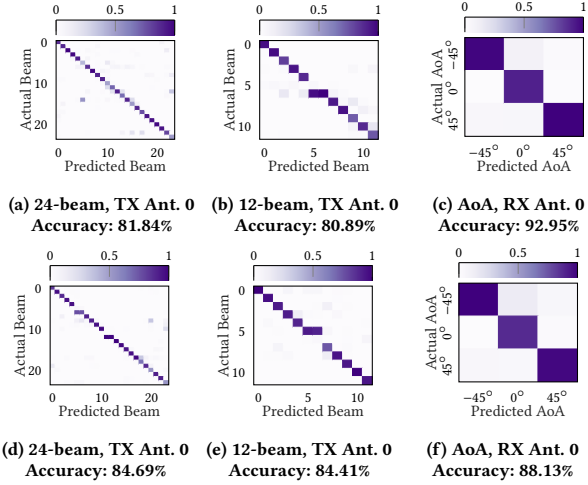

    \centering
    \captionsetup{justification=centering}
    \begin{subfigure}[t]{0.31\columnwidth}
    \centering
    \setlength\fwidth{.6\columnwidth}
    \setlength\fheight{0.5\columnwidth}
    \input{./latex_figs/diagonal_24_beam_tm_0_l_1}
    \caption{24-beam, TX Ant. 0 \\Accuracy: 81.84\%}
    \label{fig:beam_pattern_24_atn_0}
    \end{subfigure}
    \begin{subfigure}[t]{0.31\columnwidth}
    \centering
    \setlength\fwidth{.6\columnwidth}
    \setlength\fheight{0.5\columnwidth}
    \begin{tikzpicture}
\pgfplotsset{every tick label/.append style={font=\tiny}}

\begin{axis}[
enlargelimits=false,
colorbar,
colormap/Purples,
width=\fwidth,
height=\fheight,
at={(0\fwidth,0\fheight)},
scale only axis,
tick align=inside,
xlabel={Predicted Beam},
xmin=-0.5,
xmax=11.5,
xtick style={draw=none},
xlabel style={font=\scriptsize\color{white!15!black}},
ylabel style={font=\scriptsize\color{white!15!black}},
ylabel={Actual Beam},
ymin=-0.5,
ymax=11.5,
xlabel shift=-5pt,
ylabel shift=-5pt,
ytick style={draw=none},
axis background/.style={fill=white},
colorbar horizontal,
colorbar style={
at={(0,1.05)},               
anchor=below south west,    
width=\pgfkeysvalueof{/pgfplots/parent axis width},
xtick={0, 0.5, 1},
xmin=0,
xmax=1,
axis x line*=top,
xticklabel shift=-1pt,
point meta min=0,
point meta max=1,
},
colorbar/width=2mm,
]
\addplot [matrix plot,point meta=explicit]
 coordinates {
(0,0) [0.9535] (0,1) [0.004833333333333334] (0,2) [0.0009166666666666666] (0,3) [0.013166666666666667] (0,4) [0.0] (0,5) [0.019166666666666665] (0,6) [0.0] (0,7) [0.0005833333333333334] (0,8) [0.0] (0,9) [0.0020833333333333333] (0,10) [0.0] (0,11) [0.00575] 

(1,0) [0.0] (1,1) [0.9594166666666667] (1,2) [0.0008333333333333334] (1,3) [0.0] (1,4) [0.0] (1,5) [0.03758333333333334] (1,6) [0.0020833333333333333] (1,7) [0.0] (1,8) [0.0] (1,9) [0.0] (1,10) [0.0] (1,11) [8.333333333333333e-05] 

(2,0) [0.0] (2,1) [0.00025] (2,2) [0.7991666666666667] (2,3) [0.021166666666666667] (2,4) [0.0] (2,5) [0.0195] (2,6) [0.15491666666666667] (2,7) [0.0] (2,8) [0.0] (2,9) [0.0] (2,10) [0.0] (2,11) [0.005] 

(3,0) [0.003916666666666667] (3,1) [0.0005833333333333335] (3,2) [0.004333333333333334] (3,3) [0.9325000000000001] (3,4) [0.006750000000000001] (3,5) [0.04266666666666667] (3,6) [0.001166666666666667] (3,7) [0.0] (3,8) [0.00041666666666666675] (3,9) [8.333333333333334e-05] (3,10) [0.0] (3,11) [0.007583333333333334] 

(4,0) [0.0] (4,1) [0.0] (4,2) [0.0] (4,3) [0.0] (4,4) [0.8665] (4,5) [0.0] (4,6) [0.1335] (4,7) [0.0] (4,8) [0.0] (4,9) [0.0] (4,10) [0.0] (4,11) [0.0] 

(5,0) [0.0] (5,1) [0.0] (5,2) [0.003] (5,3) [0.0006666666666666666] (5,4) [0.0035833333333333333] (5,5) [0.03716666666666667] (5,6) [0.9555] (5,7) [0.0] (5,8) [0.0] (5,9) [0.0] (5,10) [0.0] (5,11) [8.333333333333333e-05] 

(6,0) [0.0] (6,1) [0.0] (6,2) [0.00275] (6,3) [8.333333333333333e-05] (6,4) [0.0026666666666666666] (6,5) [8.333333333333333e-05] (6,6) [0.9936666666666667] (6,7) [0.0] (6,8) [0.00075] (6,9) [0.0] (6,10) [0.0] (6,11) [0.0] 

(7,0) [0.0] (7,1) [0.0005] (7,2) [0.002] (7,3) [0.0] (7,4) [0.0008333333333333334] (7,5) [0.01225] (7,6) [0.06816666666666667] (7,7) [0.9155833333333333] (7,8) [0.0005833333333333334] (7,9) [0.0] (7,10) [8.333333333333333e-05] (7,11) [0.0] 

(8,0) [0.0003333333333333333] (8,1) [0.0] (8,2) [0.0023333333333333335] (8,3) [0.00016666666666666666] (8,4) [0.0020833333333333333] (8,5) [0.0] (8,6) [0.019166666666666665] (8,7) [0.034083333333333334] (8,8) [0.9390833333333334] (8,9) [0.0] (8,10) [0.0018333333333333333] (8,11) [0.0009166666666666666] 

(9,0) [0.0] (9,1) [0.0] (9,2) [0.0023333333333333335] (9,3) [0.039] (9,4) [0.0] (9,5) [0.002916666666666667] (9,6) [0.0] (9,7) [0.14641666666666667] (9,8) [0.005] (9,9) [0.7145833333333333] (9,10) [0.07391666666666667] (9,11) [0.015833333333333335] 

(10,0) [0.0] (10,1) [0.002] (10,2) [0.0] (10,3) [0.0] (10,4) [0.0] (10,5) [0.0] (10,6) [8.333333333333333e-05] (10,7) [0.018666666666666668] (10,8) [0.096] (10,9) [0.00325] (10,10) [0.86625] (10,11) [0.01375] 

(11,0) [0.0] (11,1) [8.333333333333333e-05] (11,2) [8.333333333333333e-05] (11,3) [0.059583333333333335] (11,4) [0.00016666666666666666] (11,5) [0.0005833333333333334] (11,6) [0.0005] (11,7) [0.0003333333333333333] (11,8) [0.041666666666666664] (11,9) [0.00016666666666666666] (11,10) [0.1675] (11,11) [0.7293333333333333] 

};
\end{axis}
\end{tikzpicture}
    \caption{12-beam, TX Ant. 0\\Accuracy: 80.89\%}
    \label{fig:beam_pattern_12_ant_0}
    \end{subfigure}
    \begin{subfigure}[t]{0.31\columnwidth}
    \centering
    \setlength\fwidth{.6\columnwidth}
    \setlength\fheight{0.5\columnwidth}
    \begin{tikzpicture}
\pgfplotsset{every tick label/.append style={font=\tiny}}

\begin{axis}[
enlargelimits=false,
colorbar,
colormap/Purples,
width=\fwidth,
height=\fheight,
at={(0\fwidth,0\fheight)},
scale only axis,
tick align=inside,
xlabel={Predicted AoA},
xmin=-0.5,
xmax=2.5,
xtick style={draw=none},
xlabel style={font=\scriptsize\color{white!15!black}},
ylabel style={font=\scriptsize\color{white!15!black}},
ylabel={Actual AoA},
ymin=-0.5,
ymax=2.5,
xlabel shift=-5pt,
ylabel shift=-5pt,
xtick={0,1,2},
xticklabels={$-45^\circ$, $0^\circ$, $45^\circ$},
ytick={0,1,2},
yticklabels={$-45^\circ$, $0^\circ$, $45^\circ$},
yticklabel style={rotate=90, anchor=center},
yticklabel shift=5pt,
ytick style={draw=none},
axis background/.style={fill=white},
colorbar horizontal,
colorbar style={
at={(0,1.05)},               
anchor=below south west,    
width=\pgfkeysvalueof{/pgfplots/parent axis width},
xtick={0, 0.5, 1},
xmin=0,
xmax=1,
axis x line*=top,
xticklabel shift=-1pt,
point meta min=0,
point meta max=1,
},
colorbar/width=2mm,
]
\addplot [matrix plot,point meta=explicit]
 coordinates {
(0,0) [0.9537777777777778] (0,1) [0.002944444444444445] (0,2) [0.04327777777777778] 

(1,0) [0.08722222222222223] (1,1) [0.8678888888888889] (1,2) [0.04488888888888889] 

(2,0) [0.024055555555555556] (2,1) [0.008833333333333334] (2,2) [0.9671111111111111] 

};
\end{axis}
\end{tikzpicture}
    \caption{AoA, RX Ant. 0\\Accuracy: 92.95\%}
    \label{fig:beam_pattern_aoa_ant_0}
    \end{subfigure}
    \begin{subfigure}[t]{0.31\columnwidth}
    \centering
    \setlength\fwidth{.6\columnwidth}
    \setlength\fheight{0.5\columnwidth}
    \input{./latex_figs/obstacle_24_beam_tm_0_l_1}
    \caption{24-beam, TX Ant. 0\\Accuracy: 84.69\%}
    \label{fig:beam_pattern_24_ant_0Obst}
    \end{subfigure}
    \begin{subfigure}[t]{0.31\columnwidth}
    \centering
    \setlength\fwidth{.6\columnwidth}
    \setlength\fheight{0.5\columnwidth}
    \begin{tikzpicture}
\pgfplotsset{every tick label/.append style={font=\tiny}}

\begin{axis}[
enlargelimits=false,
colorbar,
colormap/Purples,
width=\fwidth,
height=\fheight,
at={(0\fwidth,0\fheight)},
scale only axis,
tick align=inside,
xlabel={Predicted Beam},
xmin=-0.5,
xmax=11.5,
xtick style={draw=none},
xlabel style={font=\scriptsize\color{white!15!black}},
ylabel style={font=\scriptsize\color{white!15!black}},
ylabel={Actual Beam},
ymin=-0.5,
ymax=11.5,
xlabel shift=-5pt,
ylabel shift=-5pt,
ytick style={draw=none},
axis background/.style={fill=white},
colorbar horizontal,
colorbar style={
at={(0,1.05)},               
anchor=below south west,    
width=\pgfkeysvalueof{/pgfplots/parent axis width},
xtick={0, 0.5, 1},
xmin=0,
xmax=1,
axis x line*=top,
xticklabel shift=-1pt,
point meta min=0,
point meta max=1,
},
colorbar/width=2mm,
]
\addplot [matrix plot,point meta=explicit]
 coordinates {
(0,0) [0.9985833333333334] (0,1) [0.0013333333333333333] (0,2) [0.0] (0,3) [0.0] (0,4) [0.0] (0,5) [0.0] (0,6) [0.0] (0,7) [0.0] (0,8) [0.0] (0,9) [8.333333333333333e-05] (0,10) [0.0] (0,11) [0.0] 

(1,0) [0.002] (1,1) [0.9534166666666667] (1,2) [0.0] (1,3) [0.0008333333333333334] (1,4) [0.029083333333333333] (1,5) [0.0011666666666666668] (1,6) [8.333333333333333e-05] (1,7) [0.0] (1,8) [0.013333333333333334] (1,9) [8.333333333333333e-05] (1,10) [0.0] (1,11) [0.0] 

(2,0) [8.333333333333334e-05] (2,1) [0.0013333333333333335] (2,2) [0.9129166666666668] (2,3) [0.00041666666666666675] (2,4) [0.0003333333333333334] (2,5) [0.008666666666666668] (2,6) [0.0] (2,7) [0.06325000000000001] (2,8) [0.009500000000000001] (2,9) [0.0034166666666666672] (2,10) [0.0] (2,11) [8.333333333333334e-05] 

(3,0) [0.05416666666666667] (3,1) [0.00016666666666666666] (3,2) [0.0013333333333333333] (3,3) [0.9098333333333334] (3,4) [0.003416666666666667] (3,5) [0.00125] (3,6) [8.333333333333333e-05] (3,7) [0.0006666666666666666] (3,8) [0.028916666666666667] (3,9) [0.00016666666666666666] (3,10) [0.0] (3,11) [0.0] 

(4,0) [0.0] (4,1) [0.0005000000000000001] (4,2) [0.0] (4,3) [0.11266666666666668] (4,4) [0.8825000000000001] (4,5) [0.0012500000000000002] (4,6) [0.0] (4,7) [0.0] (4,8) [0.0030833333333333338] (4,9) [0.0] (4,10) [0.0] (4,11) [0.0] 

(5,0) [0.04625] (5,1) [0.0] (5,2) [0.0031666666666666666] (5,3) [0.020083333333333335] (5,4) [0.0011666666666666668] (5,5) [0.9160833333333334] (5,6) [0.012833333333333334] (5,7) [0.0] (5,8) [0.0] (5,9) [0.0003333333333333333] (5,10) [0.0] (5,11) [8.333333333333333e-05] 

(6,0) [0.04608333333333334] (6,1) [0.0001666666666666667] (6,2) [0.008250000000000002] (6,3) [0.013000000000000001] (6,4) [0.004416666666666668] (6,5) [0.8912500000000001] (6,6) [0.03325000000000001] (6,7) [0.0] (6,8) [0.00041666666666666675] (6,9) [0.0030833333333333338] (6,10) [0.0] (6,11) [8.333333333333334e-05] 

(7,0) [0.002] (7,1) [0.001] (7,2) [0.09683333333333333] (7,3) [0.0005833333333333334] (7,4) [8.333333333333333e-05] (7,5) [0.0405] (7,6) [0.0018333333333333333] (7,7) [0.7331666666666666] (7,8) [0.12066666666666667] (7,9) [0.0031666666666666666] (7,10) [0.0] (7,11) [0.00016666666666666666] 

(8,0) [0.014166666666666668] (8,1) [0.0061666666666666675] (8,2) [0.0] (8,3) [0.03308333333333334] (8,4) [0.0020000000000000005] (8,5) [0.0] (8,6) [0.00041666666666666675] (8,7) [0.004750000000000001] (8,8) [0.9378333333333334] (8,9) [0.001416666666666667] (8,10) [8.333333333333334e-05] (8,11) [8.333333333333334e-05] 

(9,0) [0.01025] (9,1) [0.0025] (9,2) [0.01025] (9,3) [0.0] (9,4) [0.0004166666666666667] (9,5) [0.010916666666666667] (9,6) [0.006166666666666667] (9,7) [0.016416666666666666] (9,8) [0.010833333333333334] (9,9) [0.93] (9,10) [0.00125] (9,11) [0.001] 

(10,0) [0.0] (10,1) [0.0] (10,2) [0.0] (10,3) [0.0] (10,4) [0.00075] (10,5) [0.0003333333333333333] (10,6) [0.0010833333333333333] (10,7) [0.0] (10,8) [0.0006666666666666666] (10,9) [0.0004166666666666667] (10,10) [0.9929166666666667] (10,11) [0.003833333333333333] 

(11,0) [0.0] (11,1) [0.0] (11,2) [0.0] (11,3) [0.0] (11,4) [0.008166666666666666] (11,5) [0.00016666666666666666] (11,6) [0.05825] (11,7) [0.0] (11,8) [0.0] (11,9) [0.0] (11,10) [0.004583333333333333] (11,11) [0.9288333333333333] 

};
\end{axis}
\end{tikzpicture}
    \caption{12-beam, TX Ant. 0\\Accuracy: 84.41\%}
    \label{fig:beam_pattern_12_ant_0Obst}
    \end{subfigure}
    \begin{subfigure}[t]{0.31\columnwidth}
    \centering
    \setlength\fwidth{.6\columnwidth}
    \setlength\fheight{0.5\columnwidth}
    \begin{tikzpicture}
\pgfplotsset{every tick label/.append style={font=\tiny}}

\begin{axis}[
enlargelimits=false,
colorbar,
colormap/Purples,
width=\fwidth,
height=\fheight,
at={(0\fwidth,0\fheight)},
scale only axis,
tick align=inside,
xlabel={Predicted AoA},
xmin=-0.5,
xmax=2.5,
xtick style={draw=none},
xlabel style={font=\scriptsize\color{white!15!black}},
ylabel style={font=\scriptsize\color{white!15!black}},
ylabel={Actual AoA},
ymin=-0.5,
ymax=2.5,
xlabel shift=-5pt,
ylabel shift=-5pt,
ytick style={draw=none},
xtick={0,1,2},
xticklabels={$-45^\circ$, $0^\circ$, $45^\circ$},
ytick={0,1,2},
yticklabels={$-45^\circ$, $0^\circ$, $45^\circ$},
yticklabel style={rotate=90, anchor=center},
yticklabel shift=5pt,
axis background/.style={fill=white},
colorbar horizontal,
colorbar style={
at={(0,1.05)},               
anchor=below south west,    
width=\pgfkeysvalueof{/pgfplots/parent axis width},
xtick={0, 0.5, 1},
xmin=0,
xmax=1,
axis x line*=top,
xticklabel shift=-1pt,
point meta min=0,
point meta max=1,
},
colorbar/width=2mm,
]
\addplot [matrix plot,point meta=explicit]
 coordinates {
(0,0) [0.9286388888888889] (0,1) [0.016583333333333332] (0,2) [0.05477777777777778] 

(1,0) [0.1435] (1,1) [0.8075555555555556] (1,2) [0.04894444444444444] 

(2,0) [0.057916666666666665] (2,1) [0.034166666666666665] (2,2) [0.9079166666666667] 

};
\end{axis}
\end{tikzpicture}
    \caption{AoA, RX Ant. 0\\Accuracy: 88.13\%}
    \label{fig:beam_pattern_aoa_ant_0_obst}
    \end{subfigure}
        \setlength\abovecaptionskip{0.2cm}
        \setlength\belowcaptionskip{-.5cm}
    \caption{Diagonal and Obstacle results with the single-RF-chain testbed.}
        \label{fig:diagonal_obstacle}
\end{figure}

To analyze the impact of the scenario on DeepBeam, we show the results obtained in the Diagonal and Obstacle configurations in Figure~\ref{fig:diagonal_obstacle}. Interestingly, we see that our CNN is flexible and robust to different conditions. In the case of 24-beam, we see a significant increase in accuracy with respect to the basic configuration, while the other accuracy results are in line with what experienced in the basic configuration.

\vspace{-.2cm}
\subsection{Training and Testing on Different Devices}
\label{sec:tota}
\vspace{-.1cm}

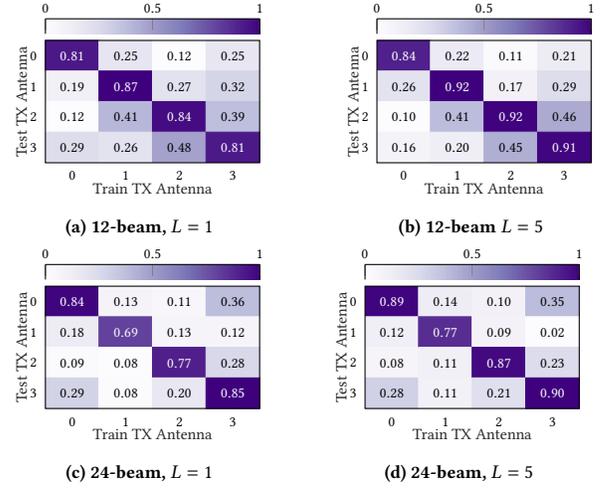
\begin{figure}[t]
    \centering
    \begin{subfigure}[t]{0.48\columnwidth}
    \centering
    \setlength\fwidth{.7\columnwidth}
    \setlength\fheight{0.4\columnwidth}
    \begin{tikzpicture}
\pgfplotsset{every tick label/.append style={font=\tiny}}

\begin{axis}[
enlargelimits=false,
colorbar,
colormap/Purples,
width=\fwidth,
height=\fheight,
at={(0\fwidth,0\fheight)},
scale only axis,
tick align=inside,
xlabel={Train TX Antenna},
xmin=-0.5,
xmax=3.5,
xtick style={draw=none},
xlabel style={font=\scriptsize\color{white!15!black}},
ylabel style={font=\scriptsize\color{white!15!black}},
ylabel={Test TX Antenna},
ymin=-0.5,
ymax=3.5,
xlabel shift=-5pt,
ylabel shift=-5pt,
ytick style={draw=none},
axis background/.style={fill=white},
colorbar horizontal,
colorbar style={
at={(0,1.05)},               
anchor=below south west,    
width=\pgfkeysvalueof{/pgfplots/parent axis width},
xtick={0, 0.5, 1},
xmin=0,
xmax=1,
axis x line*=top,
xticklabel shift=-1pt,
point meta min=0,
point meta max=1,
},
nodes near coords bottom/.style={
    scatter/position=absolute,
    close to zero/.style={at={(axis cs:\pgfkeysvalueof{/data point/x},\pgfkeysvalueof{/data point/y})}, xshift=0pt, yshift=-5pt, black, font=\tiny,
    /pgf/number format/.cd,
    fixed,
    fixed zerofill,
    precision=2,
    /tikz/.cd
    },
    big value/.style={at={(axis cs:\pgfkeysvalueof{/data point/x},\pgfkeysvalueof{/data point/y})}, xshift=0pt, yshift=-5pt, white, font=\tiny,
    /pgf/number format/.cd,
    fixed,
    fixed zerofill,
    precision=2,
    /tikz/.cd
    },
    every node near coord/.style={
      check for zero/.code={%
        \pgfmathfloatifflags{\pgfplotspointmeta}{0}{%
            \pgfkeys{/tikz/coordinate}%
        }{%
            \begingroup
            \pgfkeys{/pgf/fpu}%
            \pgfmathparse{\pgfplotspointmeta<#1}%
            \global\let\result=\pgfmathresult
            \endgroup
            %
            %
            \pgfmathfloatcreate{1}{1.0}{0}%
            \let\ONE=\pgfmathresult
            \ifx\result\ONE
                \pgfkeysalso{/pgfplots/close to zero}%
            \else
                \pgfkeysalso{/pgfplots/big value}%
            \fi
        }
      },
      check for zero, 
    },%
},%
nodes near coords bottom=0.5,
colorbar/width=2mm,
nodes near coords={\pgfmathprintnumber\pgfplotspointmeta},
]
\addplot [matrix plot,point meta=explicit]
 coordinates {
(0,0) [0.8102] (0,1) [0.1894] (0,2) [0.1190] (0,3) [0.2920] 

(1,0) [0.2513] (1,1) [0.8663] (1,2) [0.4140] (1,3) [0.2635] 

(2,0) [0.1181] (2,1) [0.2654] (2,2) [0.8401] (2,3) [0.4750] 

(3,0) [0.2478] (3,1) [0.3216] (3,2) [0.3918] (3,3) [0.8067] 
};
\end{axis}
\end{tikzpicture}

%
%



%

    \caption{12-beam, $L=1$}
    \label{fig:beam_pattern_tt_12_1}
    \end{subfigure}
    \hfill
    \begin{subfigure}[t]{0.48\columnwidth}
    \centering
    \setlength\fwidth{.7\columnwidth}
    \setlength\fheight{0.4\columnwidth}
    \begin{tikzpicture}
\pgfplotsset{every tick label/.append style={font=\tiny}}

\begin{axis}[
enlargelimits=false,
colorbar,
colormap/Purples,
width=\fwidth,
height=\fheight,
at={(0\fwidth,0\fheight)},
scale only axis,
tick align=inside,
xlabel={Train TX Antenna},
xmin=-0.5,
xmax=3.5,
xtick style={draw=none},
xlabel style={font=\scriptsize\color{white!15!black}},
ylabel style={font=\scriptsize\color{white!15!black}},
ylabel={Test TX Antenna},
ymin=-0.5,
ymax=3.5,
xlabel shift=-5pt,
ylabel shift=-5pt,
ytick style={draw=none},
axis background/.style={fill=white},
colorbar horizontal,
colorbar style={
at={(0,1.05)},               
anchor=below south west,    
width=\pgfkeysvalueof{/pgfplots/parent axis width},
xtick={0, 0.5, 1},
xmin=0,
xmax=1,
axis x line*=top,
xticklabel shift=-1pt,
point meta min=0,
point meta max=1,
},
colorbar/width=2mm,
nodes near coords bottom/.style={
    scatter/position=absolute,
    close to zero/.style={at={(axis cs:\pgfkeysvalueof{/data point/x},\pgfkeysvalueof{/data point/y})}, xshift=0pt, yshift=-5pt, black, font=\tiny,
    /pgf/number format/.cd,
    fixed,
    fixed zerofill,
    precision=2,
    /tikz/.cd
    },
    big value/.style={at={(axis cs:\pgfkeysvalueof{/data point/x},\pgfkeysvalueof{/data point/y})}, xshift=0pt, yshift=-5pt, white, font=\tiny,
    /pgf/number format/.cd,
    fixed,
    fixed zerofill,
    precision=2,
    /tikz/.cd
    },
    every node near coord/.style={
      check for zero/.code={%
        \pgfmathfloatifflags{\pgfplotspointmeta}{0}{%
            \pgfkeys{/tikz/coordinate}%
        }{%
            \begingroup
            \pgfkeys{/pgf/fpu}%
            \pgfmathparse{\pgfplotspointmeta<#1}%
            \global\let\result=\pgfmathresult
            \endgroup
            %
            %
            \pgfmathfloatcreate{1}{1.0}{0}%
            \let\ONE=\pgfmathresult
            \ifx\result\ONE
                \pgfkeysalso{/pgfplots/close to zero}%
            \else
                \pgfkeysalso{/pgfplots/big value}%
            \fi
        }
      },
      check for zero, 
    },%
},%
nodes near coords bottom=0.5,
nodes near coords={\pgfmathprintnumber\pgfplotspointmeta},
]
\addplot [matrix plot,point meta=explicit]
 coordinates {
(0,0) [0.8402] (0,1) [0.2591] (0,2) [0.0974] (0,3) [0.1577] 

(1,0) [0.2156] (1,1) [0.9243] (1,2) [0.4093] (1,3) [0.2006] 

(2,0) [0.1102] (2,1) [0.1711] (2,2) [0.9189] (2,3) [0.4520] 

(3,0) [0.2084] (3,1) [0.2889] (3,2) [0.4645] (3,3) [0.9127] 
};
\end{axis}
\end{tikzpicture}

%
    \caption{12-beam $L=5$}
    \label{fig:beam_pattern_tt_12_5}
    \end{subfigure}
    
    \begin{subfigure}[t]{0.48\columnwidth}
    \centering
    \setlength\fwidth{.7\columnwidth}
    \setlength\fheight{0.4\columnwidth}
    \begin{tikzpicture}
\pgfplotsset{every tick label/.append style={font=\tiny}}

\begin{axis}[
enlargelimits=false,
colorbar,
colormap/Purples,
width=\fwidth,
height=\fheight,
at={(0\fwidth,0\fheight)},
scale only axis,
tick align=inside,
xlabel={Train TX Antenna},
xmin=-0.5,
xmax=3.5,
xtick style={draw=none},
xlabel style={font=\scriptsize\color{white!15!black}},
ylabel style={font=\scriptsize\color{white!15!black}},
ylabel={Test TX Antenna},
ymin=-0.5,
ymax=3.5,
xlabel shift=-5pt,
ylabel shift=-5pt,
ytick style={draw=none},
axis background/.style={fill=white},
colorbar horizontal,
colorbar style={
at={(0,1.05)},               
anchor=below south west,    
width=\pgfkeysvalueof{/pgfplots/parent axis width},
xtick={0, 0.5, 1},
xmin=0,
xmax=1,
axis x line*=top,
xticklabel shift=-1pt,
point meta min=0,
point meta max=1,
},
colorbar/width=2mm,
nodes near coords bottom/.style={
    scatter/position=absolute,
    close to zero/.style={at={(axis cs:\pgfkeysvalueof{/data point/x},\pgfkeysvalueof{/data point/y})}, xshift=0pt, yshift=-5pt, black, font=\tiny,
    /pgf/number format/.cd,
    fixed,
    fixed zerofill,
    precision=2,
    /tikz/.cd
    },
    big value/.style={at={(axis cs:\pgfkeysvalueof{/data point/x},\pgfkeysvalueof{/data point/y})}, xshift=0pt, yshift=-5pt, white, font=\tiny,
    /pgf/number format/.cd,
    fixed,
    fixed zerofill,
    precision=2,
    /tikz/.cd
    },
    every node near coord/.style={
      check for zero/.code={%
        \pgfmathfloatifflags{\pgfplotspointmeta}{0}{%
            \pgfkeys{/tikz/coordinate}%
        }{%
            \begingroup
            \pgfkeys{/pgf/fpu}%
            \pgfmathparse{\pgfplotspointmeta<#1}%
            \global\let\result=\pgfmathresult
            \endgroup
            %
            %
            \pgfmathfloatcreate{1}{1.0}{0}%
            \let\ONE=\pgfmathresult
            \ifx\result\ONE
                \pgfkeysalso{/pgfplots/close to zero}%
            \else
                \pgfkeysalso{/pgfplots/big value}%
            \fi
        }
      },
      check for zero, 
    },%
},%
nodes near coords bottom=0.5,
nodes near coords={\pgfmathprintnumber\pgfplotspointmeta},
]
\addplot [matrix plot,point meta=explicit]
 coordinates {
(0,0) [0.8369] (0,1) [0.1804] (0,2) [0.0879] (0,3) [0.2920] 

(1,0) [0.1297] (1,1) [0.6877] (1,2) [0.0751] (1,3) [0.0821] 

(2,0) [0.1061] (2,1) [0.1258] (2,2) [0.7696] (2,3) [0.2013] 

(3,0) [0.3617] (3,1) [0.1203] (3,2) [0.2751] (3,3) [0.8465] 
};
\end{axis}
\end{tikzpicture}

%
%



%

    \caption{24-beam, $L=1$}
    \label{fig:beam_pattern_tt_24_1}
    \end{subfigure}
    \hfill
    \begin{subfigure}[t]{0.48\columnwidth}
    \centering
    \setlength\fwidth{.7\columnwidth}
    \setlength\fheight{0.4\columnwidth}
    \begin{tikzpicture}
\pgfplotsset{every tick label/.append style={font=\tiny}}

\begin{axis}[
enlargelimits=false,
colorbar,
colormap/Purples,
width=\fwidth,
height=\fheight,
at={(0\fwidth,0\fheight)},
scale only axis,
tick align=inside,
xlabel={Train TX Antenna},
xmin=-0.5,
xmax=3.5,
xtick style={draw=none},
xlabel style={font=\scriptsize\color{white!15!black}},
ylabel style={font=\scriptsize\color{white!15!black}},
ylabel={Test TX Antenna},
ymin=-0.5,
ymax=3.5,
xlabel shift=-5pt,
ylabel shift=-5pt,
ytick style={draw=none},
axis background/.style={fill=white},
colorbar horizontal,
colorbar style={
at={(0,1.05)},               
anchor=below south west,    
width=\pgfkeysvalueof{/pgfplots/parent axis width},
xtick={0, 0.5, 1},
xmin=0,
xmax=1,
axis x line*=top,
xticklabel shift=-1pt,
point meta min=0,
point meta max=1,
},
colorbar/width=2mm,
nodes near coords bottom/.style={
    scatter/position=absolute,
    close to zero/.style={at={(axis cs:\pgfkeysvalueof{/data point/x},\pgfkeysvalueof{/data point/y})}, xshift=0pt, yshift=-5pt, black, font=\tiny,
    /pgf/number format/.cd,
    fixed,
    fixed zerofill,
    precision=2,
    /tikz/.cd
    },
    big value/.style={at={(axis cs:\pgfkeysvalueof{/data point/x},\pgfkeysvalueof{/data point/y})}, xshift=0pt, yshift=-5pt, white, font=\tiny,
    /pgf/number format/.cd,
    fixed,
    fixed zerofill,
    precision=2,
    /tikz/.cd
    },
    every node near coord/.style={
      check for zero/.code={%
        \pgfmathfloatifflags{\pgfplotspointmeta}{0}{%
            \pgfkeys{/tikz/coordinate}%
        }{%
            \begingroup
            \pgfkeys{/pgf/fpu}%
            \pgfmathparse{\pgfplotspointmeta<#1}%
            \global\let\result=\pgfmathresult
            \endgroup
            %
            %
            \pgfmathfloatcreate{1}{1.0}{0}%
            \let\ONE=\pgfmathresult
            \ifx\result\ONE
                \pgfkeysalso{/pgfplots/close to zero}%
            \else
                \pgfkeysalso{/pgfplots/big value}%
            \fi
        }
      },
      check for zero, 
    },%
},%
nodes near coords bottom=0.5,
nodes near coords={\pgfmathprintnumber\pgfplotspointmeta},
]
\addplot [matrix plot,point meta=explicit]
 coordinates {
(0,0) [0.8887] (0,1) [0.1229] (0,2) [0.0751] (0,3) [0.2788]

(1,0) [0.1360] (1,1) [0.7746] (1,2) [0.1096] (1,3) [0.1055] 

(2,0) [0.1021] (2,1) [0.0926] (2,2) [0.8739] (2,3) [0.2132] 

(3,0) [0.3517] (3,1) [0.0237] (3,2) [0.2308] (3,3) [0.8985] 
};
\end{axis}
\end{tikzpicture}

%
    \caption{24-beam, $L=5$}
    \label{fig:beam_pattern_tt_24_5}
    \end{subfigure}
        \setlength\abovecaptionskip{0.2cm}
    \setlength\belowcaptionskip{-.5cm}
    \caption{Train One Test Another, 24-beam and 12-beam codebook with the single-RF-chain testbed.}
        \label{fig:t1ta_12_24}
\end{figure}

To understand whether the features learned by the CNN are related to the single antenna under consideration or generalize to multiple antennas, Figure \ref{fig:t1ta_12_24} shows the accuracy results obtained by training on one antenna and testing on another (TOTA), with both codebooks and all four phased arrays, with different values of $L$. Therefore, the main diagonal shows the results for train and test with the same antenna (TTSA). The first insight revealed by Figure \ref{fig:t1ta_12_24} is that the features learned by the CNN are a mixture of antenna-based and antenna-independent, since (i) the accuracy decreases when a CNN is tested on a dataset collected for a different antenna, but (ii) the accuracy does not plummet to random prediction. 

Indeed, while the average TTSA accuracy is 83.08\%, the average TOTA accuracy is 27.90\%, which is more than 3x the random guess (1/12) in the 12-beam case when $L=1$. Interestingly enough, we observe that while the average TTSA accuracy increase to 89.90\% when $L=5$, the average TOTA accuracy slightly decreases to 25.29\%.  This can be explained by the fact that a larger model will be more prone to overfitting. In this case, a smaller model will lead to less accuracy but more generalization. A similar effect is observed in the 24-beam codebook, where the TTSA increases from 78.51\% to 85.91\% between $L=1$ and $L=5$, but the TOTA slightly decreases from 16.97\% to 15.35\%.

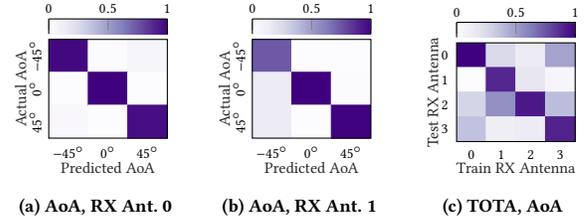
\begin{figure}[!h]
    \centering
    \vspace{-.3cm}
    \begin{subfigure}[t]{0.31\columnwidth}
    \centering
    \setlength\fwidth{.6\columnwidth}
    \setlength\fheight{0.5\columnwidth}
    \begin{tikzpicture}
\pgfplotsset{every tick label/.append style={font=\tiny}}

\begin{axis}[
enlargelimits=false,
colorbar,
colormap/Purples,
width=\fwidth,
height=\fheight,
at={(0\fwidth,0\fheight)},
scale only axis,
tick align=inside,
xlabel={Predicted AoA},
xmin=-0.5,
xmax=2.5,
xtick style={draw=none},
xtick={0,1,2},
xticklabels={$-45^\circ$, $0^\circ$, $45^\circ$},
ytick={0,1,2},
yticklabels={$-45^\circ$, $0^\circ$, $45^\circ$},
yticklabel style={rotate=90, anchor=center},
yticklabel shift=5pt,
xlabel style={font=\scriptsize\color{white!15!black}},
ylabel style={font=\scriptsize\color{white!15!black}},
ylabel={Actual AoA},
ymin=-0.5,
ymax=2.5,
xlabel shift=-5pt,
ylabel shift=-5pt,
ytick style={draw=none},
axis background/.style={fill=white},
colorbar horizontal,
colorbar style={
at={(0,1.05)},               
anchor=below south west,    
width=\pgfkeysvalueof{/pgfplots/parent axis width},
xtick={0, 0.5, 1},
xmin=0,
xmax=1,
axis x line*=top,
xticklabel shift=-1pt,
point meta min=0,
point meta max=1,
},
colorbar/width=2mm,
]
\addplot [matrix plot,point meta=explicit]
 coordinates {
(0,0) [0.9670833333333333] (0,1) [0.00047222222222222224] (0,2) [0.03244444444444444] 

(1,0) [0.010555555555555556] (1,1) [0.9894444444444445] (1,2) [0.0] 

(2,0) [0.05455555555555556] (2,1) [0.001] (2,2) [0.9444444444444444] 

};
\end{axis}
\end{tikzpicture}
    \caption{AoA, RX Ant. 0}
    \label{fig:beam_pattern_aoa_0}
    \end{subfigure}
    \begin{subfigure}[t]{0.31\columnwidth}
    \centering
    \setlength\fwidth{.6\columnwidth}
    \setlength\fheight{0.5\columnwidth}
    \begin{tikzpicture}
\pgfplotsset{every tick label/.append style={font=\tiny}}

\begin{axis}[
enlargelimits=false,
colorbar,
colormap/Purples,
width=\fwidth,
height=\fheight,
at={(0\fwidth,0\fheight)},
scale only axis,
tick align=inside,
xlabel={Predicted AoA},
xmin=-0.5,
xmax=2.5,
xtick style={draw=none},
xlabel style={font=\scriptsize\color{white!15!black}},
ylabel style={font=\scriptsize\color{white!15!black}},
ylabel={Actual AoA},
ymin=-0.5,
ymax=2.5,
xtick={0,1,2},
xticklabels={$-45^\circ$, $0^\circ$, $45^\circ$},
ytick={0,1,2},
yticklabels={$-45^\circ$, $0^\circ$, $45^\circ$},
yticklabel style={rotate=90, anchor=center},
yticklabel shift=5pt,
xlabel shift=-5pt,
ylabel shift=-5pt,
ytick style={draw=none},
axis background/.style={fill=white},
colorbar horizontal,
colorbar style={
at={(0,1.05)},               
anchor=below south west,    
width=\pgfkeysvalueof{/pgfplots/parent axis width},
xtick={0, 0.5, 1},
xmin=0,
xmax=1,
axis x line*=top,
xticklabel shift=-1pt,
point meta min=0,
point meta max=1,
},
colorbar/width=2mm,
]
\addplot [matrix plot,point meta=explicit]
 coordinates {
(0,0) [0.7248611111111111] (0,1) [0.1396111111111111] (0,2) [0.13552777777777777] 

(1,0) [0.001027777777777778] (1,1) [0.9954166666666667] (1,2) [0.003555555555555556] 

(2,0) [0.0018055555555555555] (2,1) [0.007555555555555556] (2,2) [0.9906388888888888] 

};
\end{axis}
\end{tikzpicture}
    \caption{AoA, RX Ant. 1}
    \label{fig:beam_pattern_aoa_1}
    \end{subfigure}
        \begin{subfigure}[t]{0.31\columnwidth}
    \centering
    \setlength\fwidth{.6\columnwidth}
    \setlength\fheight{0.5\columnwidth}
    \begin{tikzpicture}
\pgfplotsset{every tick label/.append style={font=\tiny}}

\begin{axis}[
enlargelimits=false,
colorbar,
colormap/Purples,
width=\fwidth,
height=\fheight,
at={(0\fwidth,0\fheight)},
scale only axis,
tick align=inside,
xlabel={Train RX Antenna},
xmin=-0.5,
xmax=3.5,
xtick style={draw=none},
xlabel style={font=\scriptsize\color{white!15!black}},
ylabel style={font=\scriptsize\color{white!15!black}},
ylabel={Test RX Antenna},
ymin=-0.5,
ymax=3.5,
xlabel shift=-5pt,
ylabel shift=-5pt,
ytick style={draw=none},
axis background/.style={fill=white},
colorbar horizontal,
colorbar style={
at={(0,1.05)},               
anchor=below south west,    
width=\pgfkeysvalueof{/pgfplots/parent axis width},
xtick={0, 0.5, 1},
xmin=0,
xmax=1,
axis x line*=top,
xticklabel shift=-1pt,
point meta min=0,
point meta max=1,
},
colorbar/width=2mm,
]
\addplot [matrix plot,point meta=explicit]
 coordinates {
(0,0) [96.69] (0,1) [45.12] (0,2) [59.15] (0,3) [63.36] 

(1,0) [58.11] (1,1) [90.36] (1,2) [73.46] (1,3) [51.33] 

(2,0) [52.32] (2,1) [53.95] (2,2) [92.79] (2,3) [52.13] 

(3,0) [69.45] (3,1) [48.05] (3,2) [64.42] (3,3) [91.18] 
};
\end{axis}
\end{tikzpicture}


%
%
    \caption{TOTA, AoA}
    \label{fig:beam_pattern_aoa_tota}
    \end{subfigure}
        \setlength\abovecaptionskip{0.2cm}
    \setlength\belowcaptionskip{-.3cm}
    \caption{AoA Results with the single-RF-chain testbed.}
        \label{fig:aoa}
\end{figure}

Figure \ref{fig:aoa} shows the results obtained after training our CNN to detect the \gls{aoa} of an incoming beam, for two different antennas. We also report the TOTA results in Figure \ref{fig:aoa}(c). Figure \ref{fig:aoa} indicates that the CNN obtains a very high accuracy of more than 90\%. As before, the TOTA results (57.57\% in this case) show that the learned features are a mix of antenna-dependent and independent features.

\begin{figure}[b]
    \centering
    \vspace{-.3cm}
    \captionsetup{justification=centering}
    \begin{subfigure}[t]{0.31\columnwidth}
    \centering
    \setlength\fwidth{.6\columnwidth}
    \setlength\fheight{0.5\columnwidth}
    \begin{tikzpicture}
\pgfplotsset{every tick label/.append style={font=\tiny}}

\begin{axis}[
enlargelimits=false,
colorbar,
colormap/Purples,
width=\fwidth,
height=\fheight,
at={(0\fwidth,0\fheight)},
scale only axis,
tick align=inside,
xlabel={Predicted beam},
xmin=-0.5,
xmax=11.5,
xtick style={draw=none},
xlabel style={font=\scriptsize\color{white!15!black}},
ylabel style={font=\scriptsize\color{white!15!black}},
ylabel={Actual beam},
ymin=-0.5,
ymax=11.5,
xlabel shift=-5pt,
ylabel shift=-5pt,
ytick style={draw=none},
axis background/.style={fill=white},
colorbar horizontal,
colorbar style={
at={(0,1.05)},               
anchor=below south west,    
width=\pgfkeysvalueof{/pgfplots/parent axis width},
xtick={0, 0.5, 1},
xmin=0,
xmax=1,
axis x line*=top,
xticklabel shift=-1pt,
point meta min=0,
point meta max=1,
},
colorbar/width=2mm,
]
\addplot [matrix plot,point meta=explicit]
 coordinates {
(0,0) [0.6120833333333333] (0,1) [0.18725] (0,2) [0.0225] (0,3) [0.0] (0,4) [0.002] (0,5) [0.0009166666666666666] (0,6) [0.0008333333333333334] (0,7) [0.06433333333333334] (0,8) [0.09275] (0,9) [0.015666666666666666] (0,10) [0.0016666666666666668] (0,11) [0.0] 

(1,0) [0.0071666666666666675] (1,1) [0.9349166666666667] (1,2) [0.0045000000000000005] (1,3) [0.0032500000000000003] (1,4) [0.006500000000000001] (1,5) [0.004250000000000001] (1,6) [0.018250000000000002] (1,7) [0.005666666666666668] (1,8) [0.004666666666666668] (1,9) [0.00041666666666666675] (1,10) [0.010333333333333335] (1,11) [8.333333333333334e-05] 

(2,0) [0.016] (2,1) [0.14275] (2,2) [0.5323333333333333] (2,3) [0.006583333333333333] (2,4) [0.0565] (2,5) [0.05675] (2,6) [0.051583333333333335] (2,7) [0.0165] (2,8) [0.10375] (2,9) [0.005333333333333333] (2,10) [0.01175] (2,11) [0.00016666666666666666] 

(3,0) [0.00041666666666666675] (3,1) [0.059000000000000004] (3,2) [0.07183333333333335] (3,3) [0.8570833333333334] (3,4) [0.0051666666666666675] (3,5) [0.00025000000000000006] (3,6) [0.001416666666666667] (3,7) [0.001166666666666667] (3,8) [0.0029166666666666672] (3,9) [0.0] (3,10) [0.0005000000000000001] (3,11) [0.00025000000000000006] 

(4,0) [0.0] (4,1) [0.03858333333333333] (4,2) [0.004583333333333333] (4,3) [0.0485] (4,4) [0.8555833333333334] (4,5) [0.017416666666666667] (4,6) [0.017333333333333333] (4,7) [0.001] (4,8) [0.017] (4,9) [0.0] (4,10) [0.0] (4,11) [0.0] 

(5,0) [0.001] (5,1) [0.09616666666666666] (5,2) [0.029416666666666667] (5,3) [0.023333333333333334] (5,4) [0.16675] (5,5) [0.06708333333333333] (5,6) [0.52325] (5,7) [0.0015] (5,8) [0.05333333333333334] (5,9) [0.0010833333333333333] (5,10) [0.035916666666666666] (5,11) [0.0011666666666666668] 

(6,0) [0.0005000000000000001] (6,1) [0.10808333333333336] (6,2) [0.04250000000000001] (6,3) [0.01941666666666667] (6,4) [0.09558333333333337] (6,5) [0.05166666666666668] (6,6) [0.6160000000000001] (6,7) [0.0075000000000000015] (6,8) [0.04683333333333334] (6,9) [8.333333333333334e-05] (6,10) [0.011416666666666669] (6,11) [0.0004166666666666668] 

(7,0) [0.07233333333333333] (7,1) [0.18033333333333335] (7,2) [0.06733333333333333] (7,3) [0.004166666666666667] (7,4) [0.005333333333333333] (7,5) [0.02775] (7,6) [0.018416666666666668] (7,7) [0.5503333333333333] (7,8) [0.0405] (7,9) [0.018666666666666668] (7,10) [0.014083333333333333] (7,11) [0.00075] 

(8,0) [0.0033333333333333335] (8,1) [0.14133333333333334] (8,2) [0.01225] (8,3) [0.06175] (8,4) [0.027666666666666666] (8,5) [0.017166666666666667] (8,6) [0.004083333333333333] (8,7) [0.01725] (8,8) [0.714] (8,9) [8.333333333333333e-05] (8,10) [0.001] (8,11) [8.333333333333333e-05] 

(9,0) [0.058583333333333334] (9,1) [0.13475] (9,2) [0.07425] (9,3) [0.005083333333333333] (9,4) [0.015916666666666666] (9,5) [0.017166666666666667] (9,6) [0.002916666666666667] (9,7) [0.024083333333333335] (9,8) [0.0135] (9,9) [0.5864166666666667] (9,10) [0.035583333333333335] (9,11) [0.03175] 

(10,0) [0.0020833333333333333] (10,1) [0.04733333333333333] (10,2) [0.02] (10,3) [0.01625] (10,4) [0.010583333333333333] (10,5) [0.02375] (10,6) [0.0005] (10,7) [0.0011666666666666668] (10,8) [0.00325] (10,9) [0.0075] (10,10) [0.20008333333333334] (10,11) [0.6675] 

(11,0) [0.0] (11,1) [0.0] (11,2) [0.0] (11,3) [0.0] (11,4) [0.002416666666666667] (11,5) [0.0] (11,6) [8.333333333333333e-05] (11,7) [0.0] (11,8) [0.0] (11,9) [0.0] (11,10) [0.0] (11,11) [0.9975] 

};
\end{axis}
\end{tikzpicture}
    \caption{12-beam, Accuracy: 62.69\%}
    \label{fig:beam_pattern_mixed_12_1}
    \end{subfigure}
    \begin{subfigure}[t]{0.31\columnwidth}
    \centering
    \setlength\fwidth{.6\columnwidth}
    \setlength\fheight{0.5\columnwidth}
    \input{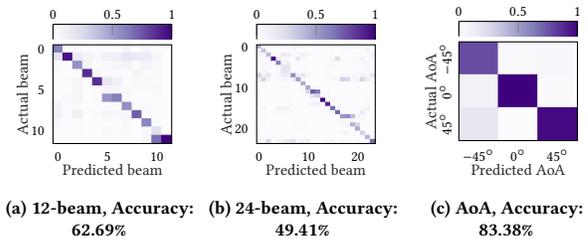}
    \caption{24-beam, Accuracy: 49.41\%}
    \label{fig:beam_pattern_mixed_24_1}
    \end{subfigure}
    \begin{subfigure}[t]{0.31\columnwidth}
    \centering
    \setlength\fwidth{.6\columnwidth}
    \setlength\fheight{0.5\columnwidth}
    \begin{tikzpicture}
\pgfplotsset{every tick label/.append style={font=\tiny}}
\begin{axis}[
enlargelimits=false,
colorbar,
colormap/Purples,
width=\fwidth,
height=\fheight,
at={(0\fwidth,0\fheight)},
scale only axis,
tick align=inside,
xlabel={Predicted AoA},
xmin=-0.5,
xmax=2.5,
xtick style={draw=none},
xlabel style={font=\scriptsize\color{white!15!black}},
ylabel style={font=\scriptsize\color{white!15!black}},
ylabel={Actual AoA},
ymin=-0.5,
ymax=2.5,
xlabel shift=-5pt,
ylabel shift=-5pt,
xtick={0,1,2},
xticklabels={$-45^\circ$, $0^\circ$, $45^\circ$},
ytick={0,1,2},
yticklabels={$-45^\circ$, $0^\circ$, $45^\circ$},
yticklabel style={rotate=90, anchor=center},
yticklabel shift=5pt,
ytick style={draw=none},
axis background/.style={fill=white},
colorbar horizontal,
colorbar style={
at={(0,1.05)},               
anchor=below south west,    
width=\pgfkeysvalueof{/pgfplots/parent axis width},
xtick={0, 0.5, 1},
xmin=0,
xmax=1,
axis x line*=top,
xticklabel shift=-1pt,
point meta min=0,
point meta max=1,
},
colorbar/width=2mm,
]
\addplot [matrix plot,point meta=explicit]
 coordinates {
(0,0) [0.70725] (0,1) [0.10783333333333334] (0,2) [0.18491666666666667] 

(1,0) [0.057972222222222224] (1,1) [0.9090277777777778] (1,2) [0.033] 

(2,0) [0.06497222222222222] (2,1) [0.04975] (2,2) [0.8852777777777778] 

};
\end{axis}
\end{tikzpicture}
    \caption{AoA, Accuracy: 83.38\%}
    \label{fig:beam_pattern_mixed_aoa}
    \end{subfigure}
            \setlength\abovecaptionskip{0.2cm}

    \caption{Mixed train/test results with the single-RF-chain testbed.}
    \label{fig:mixed}
\end{figure}

To further test the generalization capability of our CNN, we trained and tested it on a mixed dataset with waveforms coming from all 4 antennas. Figure \ref{fig:mixed} shows the obtained accuracy in all three learning problems when $L=1$, and indicates that the CNN is very effective in generalizing to different antennas, increasing the accuracy of
124\%, 191\% and 44\% in case of 24-beam, 12-beam and AoA with respect to the average TOTA accuracy experienced when trained with a single dataset. Similar results can be obtained when training on one spatial configuration (e.g., basic, diagonal, obstacle) and testing on another.

\vspace{-.2cm}
\subsection{Discussion and Possible Extensions}
\label{sec:discussion}
\vspace{-.1cm}

The results presented in Section \ref{sec:results} highlight the efficacy of our CNN-based approach for the classification of the \gls{txb} and the \gls{aoa}, demonstrating \textit{for the first time} waveform learning approaches at mmWaves and confirming the intuition described in Sec.~\ref{sec:learning_engine}. In particular, DeepBeam is robust with respect different devices, antenna architectures, input sizes, \gls{snr} levels, and deployment scenarios. We believe that this opens new and exciting research directions, fostered also by the DeepBeam dataset, toward:

\noindent $\bullet$ The development of fine-tuning solution to improve the transferability of the learning process for directional transmissions in a wider range of scenarios. The results in Sec.~\ref{sec:tota} have shown how the proposed approach is robust with respect to training and testing on different devices, and how cross-training can help DeepBeam generalizing and achieving device independence. This work can be further extended by developing protocols and algorithms to deploy DeepBeam in new scenarios and with new radios, for which re-training or fine-tuning of the DeepBeam \glspl{cnn} may be required~\cite{tajbakhsh2016convolutional}. In this regard, future work will develop automated data collection and labeling tools (e.g., with interaction between the device firmware and the DeepBeam framework) to enhance the training datasets for DeepBeam-equipped devices;

\noindent $\bullet$ The identification of the \gls{aoa} with a model-free approach, on a single-RF-chain device, and without the need for multiple sampling in space and/or time (as, for example, in~\cite{wei2017facilitating}). This paper has considered a subset of possible \gls{aoa} values for the classification, but we plan to extend the dataset and evaluation to achieve a finer detection of the \gls{aoa}.


\vspace{-.3cm}
\section{Related Work}\label{sec:rw}
\vspace{-.1cm}

Beam management is a key problem in \gls{mmwave} networks, and has attracted significant interest~\cite{Nitsche-infocom2015,va2016beam,DeDonno-ieeetwc2017,palacios2017tracking,Steinmetzer-conext2017,Loch-conext2017,sur2018towards,zhou2017beam,zhou2018following,zhou2018beam,Ghasempour-mobicom2018,Haider-mobicom2018,yang2019beam,aykin2020mamba,aykin2020efficient,giordani2018tutorial}.

One of the earliest works in the field is due to \cite{Nitsche-infocom2015}, where Nitsche \emph{et al.}  utilize eavesdropping of legacy sub-6 bands to estimate the direction for pairing nodes. Va \emph{et al.} propose a beam tracking algorithm in \cite{va2016beam}, which however assumes external \gls{aoa} and \gls{aod} estimators. The authors in \cite{DeDonno-ieeetwc2017,palacios2017tracking} propose beam training protocols that leverage hybrid analog-digital beamforming  antennas to scan multiple spatial sectors simultaneously. Steinmetzer \emph{et al.} \cite{Steinmetzer-conext2017} adapt compressive path tracking for sector selection in off-the-shelf IEEE 802.11ad devices, where the strength of received frames is used to  sweep only through a subset of probing sectors. Loch \textit{et al.} \cite{Loch-conext2017}  track both movement and rotation of 60 GHz mobile devices with a zero-overhead mechanism, where part of the preamble of each packet is transmitted using a multi-lobe beam pattern. Zhou \emph{et al.} \cite{zhou2017beam} present a model-driven approach that performs a virtual scan based on a spatial channel profile built at the receiver, which however is bootstrapped using environmental information. Sur \emph{et al.} \cite{sur2018towards} propose a beam sweeping algorithm with reduced complexity, thanks to a space-time analysis of the directional paths that reduces the number of directions to scan. Zhou \emph{et al.} present in  \cite{zhou2018following} a 3D beam sweeping along a ``cross'' around the central direction. Zhou \emph{et al.}   propose in \cite{zhou2018beam} a beam tracking mechanism to address beam misalignment between \gls{uavs}. Ghasempour \emph{et al.} \cite{Ghasempour-mobicom2018} proposed a system leveraging channel sparsity and the  knowledge of the beam  codebook to  construct a set  of  candidate  beams  for multi-stream  beam  steering. Aykin \emph{et al.} \cite{aykin2020mamba} propose a multi-armed bandit framework for beam tracking based on ACK/NACK feedback, where reinforcement learning is used to select the appropriate beams
and transmission rates. In \cite{aykin2020efficient}, the authors propose a log-time peak finding algorithm to find the best beam in a three-dimensional space.

All the above work requires some sort of coordination with the \gls{txer}, which in turn introduces overhead. Our approach in \emph{DeepBeam}, instead, is \emph{fully passive} and does not need any information exchange with the \gls{txer}. Regarding passive beam tracking, Haider \emph{et al.} proposed in \cite{Haider-mobicom2018} \emph{LiSteer}, a mechanism using external light-emitting diodes (LEDs) located on the wireless \gls{ap} to track the user's movement. However, the mechanism requires additional equipment and may not work in many circumstances (\textit{i.e.}, visible light is present, mobile device inside the pocket, and so on). Moreover, it requires traditional beam sweeping at the \gls{ap} side. In the paper, we demonstrated that our methodology is \emph{standard-agnostic} and \emph{antenna-agnostic} and can be utilized at both \gls{txer} and \gls{rxer} side.

The application of deep learning to improve the performance of wireless communications has seen a steadfast rise in the research community over the last few years. Specifically, deep learning is being used to address challenging problems such as modulation recognition, radio fingerprinting, and many others \cite{o2017introduction,OShea-ieeejstsp2018,JagannathAdHoc2019,Mao-ieeecomm2018}. The interest in this technique comes from its versatility in addressing a wide variety of wireless classification problems where an explicit mathematical model is cumbersome to obtain (\textit{e.g.}, because of the phenomenon itself or due to the scale of the classification problem). System aspects of deep learning in wireless have been also recently investigated in \cite{restuccia2019big,restuccia2020deepwierl}, which have proven its applicability to address real-time classification problems. On the other hand, machine learning in the \gls{mmwave} domain is still at its infancy, also due to the current lack of large-scale experimental databases. Existing relevant studies in the \gls{mmwave} domain use either simulations or ray tracing, which may not entirely capture the complexity of real-life propagation scenarios at \gls{mmwave} spectrum bands, and do not consider deep learning on I/Q samples~\cite{sim2018online,alkhateeb2018deep,asadi2018fml,wang2018mmwave,zhou2018deep}. To the best of our knowledge, this paper is the first to utilize experimental real-world data to address a practical deep learning problem in the \gls{mmwave} domain.


\vspace{-.4cm}
\section{Conclusions}
\vspace{-.1cm}

In this paper, we have presented \emph{DeepBeam}, a framework for beam management in \gls{mmwave} networks  that eliminates the need of beam sweeping by inferring through deep learning the direction and the \gls{aoa} of the transmitter beam. We have conducted an extensive experimental data collection campaign with two software-defined radio testbeds, and by using multiple antennas, codebooks, gains and locations. We have also implemented our learning models on \gls{fpga} to evaluate the latency performance. Results show that \emph{DeepBeam} (i) achieves accuracy of up to 96\%, 84\% and 77\% with a 5-beam, 12-beam and 24-beam codebook, respectively; (ii) reduces latency by up to 7x with respect to the 5G NR initial beam sweep in a default configuration and with a 12-beam codebook. To allow repeatability, we also share our waveform datasets and the full \emph{DeepBeam} code repository with the community.

\vspace{-.3cm}
\section*{Acknowledgements}
\vspace{-.2cm}
This work was supported in part by the US National Science Foundation
under Grants CNS-1923789 and CCF-1937500. The authors would like to thank the anonymous reviewers for their insightful comments, and Yin Sun for shepherding the paper.

\vspace{-.3cm}


\footnotesize
\bibliographystyle{ieeetr}
\bibliography{bibl}

\end{document}